\documentclass[%
 reprint,
 amsmath,amssymb,
 aps,
 prstab,
 floatfix,
]{revtex4-1}

\usepackage{graphicx}					% Graphics
\usepackage{wasysym}					% Bold PI
\usepackage{mathtools}					% Curved arrow
\usepackage[makeroom]{cancel}				% Crossed math
\newcommand{\pd}{\partial}				% Partial derivative
\newcommand{\dd}{\mathrm{d}}				% General derivative

\newcommand{\ex}{\hat{\mathbf{e}}_1}			% e_1
\newcommand{\ey}{\hat{\mathbf{e}}_2}			% e_2
\newcommand{\ez}{\hat{\mathbf{e}}_3}			% e_3
\newcommand{\tV}{\hat{\mathbf{t}}}			% t
\newcommand{\nV}{\hat{\mathbf{n}}}			% n
\newcommand{\bV}{\hat{\mathbf{b}}}			% b

\newcommand{\At}{A_\theta}			% e_x

\newcommand{\Ason}{\overline{  A   }_z    ^{(n)}}	% A_s	normal
\newcommand{\Asun}{\underline{ A   }_{\,z}^{(n)}}	% A_s	skew
\newcommand{\Aton}{\overline{  A   }_\theta    ^{(n)}}	% A_s	normal
\newcommand{\Atun}{\underline{ A   }_{\,\theta}^{(n)}}	% A_s	skew
\newcommand{\FIon}{\overline{  \Phi}      ^{(n)}}	% Phi	normal
\newcommand{\FIun}{\underline{ \Phi}      ^{(n)}}	% Phi	skew

\newcommand{\Fxon}{\overline{ F}_{\,\text{x}}^{(n)}}	% F_x	normal
\newcommand{\Fxun}{\underline{F}_{\,\text{x}}^{(n)}}	% F_x	skew
\newcommand{\Fyon}{\overline{ F}_{\,\text{y}}^{(n)}}	% F_y	normal
\newcommand{\Fyun}{\underline{F}_{\,\text{y}}^{(n)}}	% F_y	skew
\newcommand{\Fron}{\overline{ F}_{\,\rho    }^{(n)}}	% F_rho	normal
\newcommand{\Frun}{\underline{F}_{\,\rho    }^{(n)}}	% F_rho	skew

\newcommand{\An}{\mathcal{A}_n}				% A_n
\newcommand{\Bn}{\mathcal{B}_n}				% B_n

\newcommand{\Fn}{\mathcal{F}_n}				% F-functions
\newcommand{\Gn}{\mathcal{G}_n}				% G-functions

\newcommand{\Cne}{\mathcal{A}_n^{(\text{e})}}		% A_n^e
\newcommand{\Dne}{\mathcal{B}_n^{(\text{e})}}		% B_n^e
\newcommand{\Cnm}{\mathcal{A}_n^{(\text{m})}}		% A_n^m
\newcommand{\Dnm}{\mathcal{B}_n^{(\text{m})}}		% B_n^m

\newcommand{\Cno}{\overline{ C}_n    }			% Normal mltpl strength
\newcommand{\Cnu}{\underline{C}_{\,n}}			% Skew   mltpl strength

\begin{document}

%===============================================================================
\title{ On sector magnets or transverse electromagnetic fields in cylindrical 
	coordinates}
\thanks{Operated by Fermi Research Alliance, LLC under Contract
No.~De-AC02-07CH11359 with the U.S. Department of Energy.}
\author{T.~Zolkin}
\email{zolkin@fnal.gov}
\affiliation{Fermilab, PO Box 500, Batavia, IL 60510-5011}
\date{\today}
%===============================================================================

%===============================================================================
\begin{abstract}
The Laplace's equations for the scalar and vector potentials describing electric
or magnetic fields in cylindrical coordinates with translational invariance 
along azimuthal coordinate are considered.
The series of special functions which, when expanded in power series in radial
and vertical coordinates, in lowest order replicate the harmonic homogeneous
polynomials of two variables are found.
These functions are based on radial harmonics found by Edwin~M.~McMillan in his
more-than-40-years "forgotten" article, which will be discussed.
In addition to McMillan's harmonics, second family of adjoint radial harmonics
is introduced, in order to provide symmetric description between electric and
magnetic fields and to describe fields and potentials in terms of same special
functions.
Formulas to relate any transverse fields specified by the coefficients in the 
power series expansion in radial or vertical planes in cylindrical coordinates 
with the set of new functions are provided.

This result is no doubt important for potential theory while also critical
for theoretical studies, design and proper modeling of sector dipoles, combined 
function dipoles and any general sector element for accelerator physics.
All results are presented in connection with these problems.
\end{abstract}
%===============================================================================

% PACS, the Physics and Astronomy Classification Scheme.
% Use showkeys class option if keyword
                             % display desired

%===============================================================================
\maketitle
%\tableofcontents
%===============================================================================

%==============================================================================%
%==============================================================================%
%==============================================================================%
\section{Introduction}

Description of sector combined function magnets, and in general any magnet with
translational symmetry along azimuthal coordinate in cylindrical coordinates,
is very important issue, and, without any particular reference one can say that
every modern accelerator code includes such elements.
The main idea, which goes back to original 1968 K.~Brown's
paper~\cite{Brown:1982dd}, based on a solution of Laplace's equation for 
scalar potential in cylindrical coordinates using the general power series
ansatz.
Similar approach but for Laplace's equation for longitudinal component of vector
potential can be found for example in~\cite{forest1998beam}.
As one can see the approach is the same in most recent books, e.g. 
in great details in~\cite{wiedemann2015particle}.

Two major bottlenecks should be noticed.
In the first place, if one looking for a solution in a form of a series, these
series should be truncated.
In our case truncation means that potentials do not satisfy the Laplace's
equation anymore, even if symplectic integrators are used for numerical 
solution (of course potentials can ``satisfy'' the Laplace's equation up to 
desired order by keeping more and more terms in expansion).
But more importantly, the recurrence equation is undetermined.
That means in every new order of recurrence one have to assign an arbitrary 
constant, which will affect all other higher order terms.
The uncertainty leads to the fact that there is no one particular choice
of basis functions;
it make it almost impossible to compare different accelerator codes, since 
different assumptions might be used for representations of basis functions.

The indeterminacy has simple geometrical illustration.
Looking for a field with pure normal dipole component on equilibrium orbit in
lowest order, one can come up with almost arbitrary shape of magnet's north 
pole if south pole is symmetric with respect to midplane.
In the case of dipole, series can be truncated by keeping only dipole component.
For higher order multipoles in cylindrical coordinates truncation without 
violation of Laplace's equation is not possible.

Working on implementation of these magnets for Synergia, I found particular
assumptions which let me to summate series for pure electric and 
magnetic skew and normal multipoles.
Further look for symmetry in description allowed to generate full family of
solutions where no truncation is required since all series can be summated.
While discussing my results with Sergei Nagaitsev, he brought my attention,
as we found later to more-than-40-years forgotten, article by 
McMillan~\cite{McMillan:1975yw} of 1975.

Brining together his and my results I would like to present a new description
for multipole expansion in cylindrical coordinates.
Any transverse field can be expanded in terms of 
these functions and related to power series field expansion in horizontal or
vertical planes.
The new approach do not contradict with previous results but embrace it.
An ambiguity in choice of coefficients and problem of truncation are resolved.
Thus it can be employed for theoretical studies,
design and simulation of sector magnets.

%==============================================================================%
%==============================================================================%
\subsection{Article structure}

Section~\ref{sec:GE} describes general equations of motion for a particle 
in curvilinear coordinates associated with Frenet-Serret frame.
The case of transverse electromagnetic fields described in 
section~\ref{sec:TEM}.

Subsections~\ref{sec:t-R},\ref{sec:s-R} provide most general equations of 
motion for pure electric and magnetic fields.
Two further subsections~\ref{sec:RM},\ref{sec:SM} describes the expansion of 
fields in multipoles for cases with zero and constant curvatures.
The section~\ref{sec:Rec} relates new family of functions to
recurrence equations.

%==============================================================================%
%==============================================================================%
%==============================================================================%
%==============================================================================%
%==============================================================================%
\section{\label{sec:GE}General equations of motion}

%==============================================================================%
%==============================================================================%
\subsection{\label{sec:LAB}Global coordinates in Lab frame}

%------------------------------------------------------------------------------%
The Lagrangian of a relativistic particle of mass $m$ with an electric charge
$e$ in most general static electromagnetic field is given by
\[
\mathcal{L}[\mathbf{R},\dot{\mathbf{R}};t] =
	-\frac{m\,c^2}{\gamma(\mathbf{V})}
	-e\,\Phi(\mathbf{R})
	+e\left( \mathbf{V} \cdot \mathbf{A}(\mathbf{R}) \right),
\]
where $\mathbf{R}=(Q_1,Q_2,Q_3)$ is a position vector in the configuration space
of generalized coordinates spanned on three dimensional right-handed Cartesian
coordinate system
$\{\hat{\mathbf{E}}_1,\hat{\mathbf{E}}_2,\hat{\mathbf{E}}_3\}$
associated with Lab frame at the facility of a particle accelerator,
$\mathbf{V} \equiv \dot{\mathbf{R}}$ is a vector of matching generalized 
velocities where $\dot{(\ldots)}\equiv\frac{\dd}{\dd t}$ is the time derivative 
operator.
$\Phi(\mathbf{R})$ and $\mathbf{A}(\mathbf{R})$ are the electric scalar and
magnetic vector potentials respectively, and,
\[
	\gamma(\mathbf{V}) = \frac{1}{\sqrt{1-\beta(\mathbf{V})^2}}
\]
is the relativistic Lorentz factor where $\beta$ is the ratio of $V$ to the
speed of light in vacuum, $c$.

%------------------------------------------------------------------------------%
Substituting the Lagrangian into the Euler-Lagrange equations
(Lagrange's equations of the second kind)
\[
	\frac{\dd}{\dd t}
	\frac{\pd\,\mathcal{L}}{\pd \dot{\mathbf{R}}} - 
	\frac{\pd\,\mathcal{L}}{\pd      \mathbf{R} } = 0
\]
with shorthand notation
\[
	\frac{\pd}{\pd \mathbf{a}} =
	\left(
		\frac{\pd}{\pd a_1},
		\frac{\pd}{\pd a_2},
		\frac{\pd}{\pd a_3}
	\right)
\]
representing a vector of partial derivatives with respect to the indicated
variables, gives the equation of motion which is the relativistic form of the
Lorentz force
\[
	\mathbf{F} = e \left[ \mathbf{E} + (\mathbf{V}\times\mathbf{B})\right]
\]
or explicitly
\[
	\frac{\dd}{\dd t} (\gamma\,m\,\dot{Q}_i) =
	e\,(E_i + \epsilon_{ijk} \dot{Q}_j B_k)
\]
%\begin{eqnarray*}
%\frac{\dd}{\dd t} (\gamma\,m\,\dot{Q}_1) & = &
%	e\,(E_1 + \dot{Q}_2 B_3 - \dot{Q}_3 B_2),			\\
%\frac{\dd}{\dd t} (\gamma\,m\,\dot{Q}_2) & = &
%	e\,(E_2 + \dot{Q}_3 B_1 - \dot{Q}_1 B_3),			\\
%\frac{\dd}{\dd t} (\gamma\,m\,\dot{Q}_3) & = &
%	e\,(E_3 + \dot{Q}_1 B_2 - \dot{Q}_2 B_1),
%\end{eqnarray*}
where the electric and magnetic fields related to scalar electric and vector
magnetic potentials through the gradient and curl vector operators respectively
\begin{eqnarray*}
	\mathbf{E} &=& (E_1,E_2,E_3) \equiv \,\,-\,\nabla\,\Phi,	\\
	\mathbf{B} &=& (B_1,B_2,B_3) \equiv \nabla\times\mathbf{A}.	
\end{eqnarray*}

%------------------------------------------------------------------------------%
A more abstract formulation can be given in terms of Hamiltonian which describes
phase space of canonical variables $\{\mathbf{P},\mathbf{Q}\}$, where
$\mathbf{P}$ is the particle's canonical (total) momentum defined as
\[
\mathbf{P} \equiv \frac{\pd\,\mathcal{L}}{\pd \dot{\mathbf{R}}} =
	\boldsymbol{\Pi} + e\,\mathbf{A}
\]
and $\boldsymbol{\Pi} = \gamma\,m\,\mathbf{V}$ being the particle's
kinetic momentum.
The Hamiltonian might be constructed using the Legendre transformation of
$\mathcal{L}$
\begin{eqnarray*}
\mathcal{H}[\mathbf{P},\mathbf{Q};t]	& = &
	\mathbf{V}\,\mathbf{P} - \mathcal{L} =
	\sum_{i=1}^3 \dot{Q}_i P_i - \mathcal{L}			\\
					& = & c\,
	\sqrt{ m^2c^2 + \left(\mathbf{P} - e\,\mathbf{A}\right)^2}
	+ e\,\Phi.
\end{eqnarray*}
The time evolution of the system is given by Hamilton's equations
\[
\frac{\dd\,\mathbf{P}}{\dd t} =-\frac{\pd\,\mathcal{H}}{\pd\mathbf{Q}}
\qquad\text{and}\qquad
\frac{\dd\,\mathbf{Q}}{\dd t} = \frac{\pd\,\mathcal{H}}{\pd\mathbf{P}}
\]
or equivalently
\begin{eqnarray*}
\dot{\mathbf{Q}} & = &
	c\,\frac{\mathbf{P}-e\,\mathbf{A}}
		{\sqrt{m^2c^2+(\mathbf{P}-e\,\mathbf{A})^2}},     \\
\dot{\mathbf{P}} & = & e\,\left(\nabla\mathbf{A}\right)\cdot\dot{\mathbf{Q}} -
		       e\,\nabla\Phi.
\end{eqnarray*}

\begin{figure}[b!]
\includegraphics[width=0.96\linewidth]{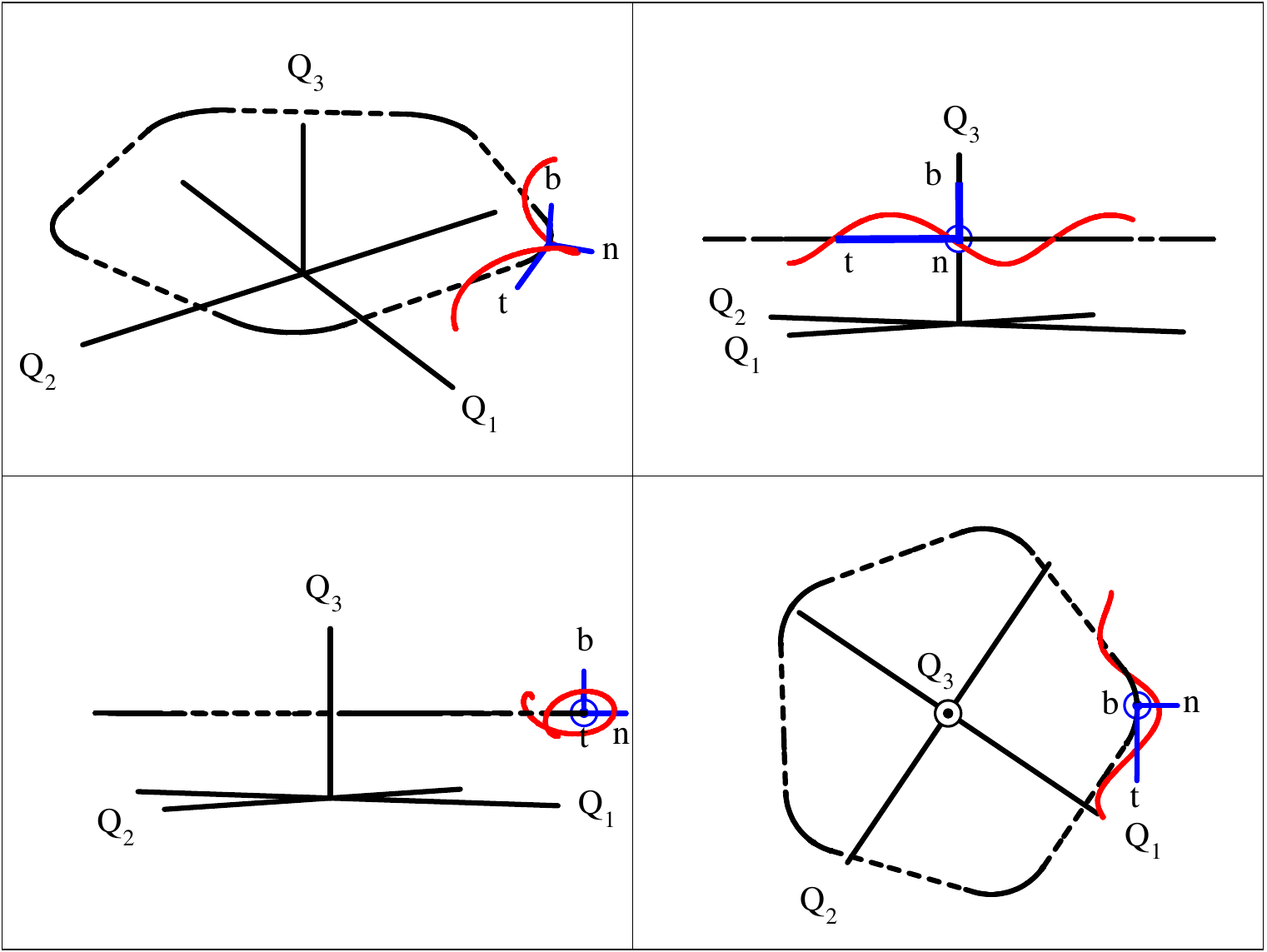}
\caption{\label{fig:LGFrames}
	Schematic plot of an equilibrium orbit for an accelerator consisting of
	five drift spaces and five $72^{\circ}$ bending magnets.
	Lab frame and local Frenet-Serret frames are shown in black and blue
	colors respectively.
	The test particle winding the equilibrium orbit shown in red.}
\end{figure}

%------------------------------------------------------------------------------%
The model of accelerator assumes the specification of a {\it reference orbit}
designed for a particle with certain equilibrium energy and assignment of beam
line elements placed along it.
In the case of a circular accelerator the closed orbit of a machine with
alignment errors in general will not coincide with reference orbit.
For most accelerator needs (except e.g. helical orbits for muon cooling)
the designed orbit is piecewise flat function, which means that it consists of a
series of curves with zero torsion;
moreover, usually, these curves are straight lines and circular arcs.
In order to better exploit the geometry of beam motion and symmetry of
electromagnetic fields we will introduce the local Frenet-Serret frame attached
to equilibrium orbit and new global coordinates associated with it
(see FIG.~\ref{fig:LGFrames}).

%------------------------------------------------------------------------------%
%------------------------------------------------------------------------------%

%==============================================================================%
%==============================================================================%
\subsection{\label{sec:SFF}Global coordinates associated with \\
	    Frenet-Serret frame}

\begin{figure}[t!]
\includegraphics[width=\linewidth]{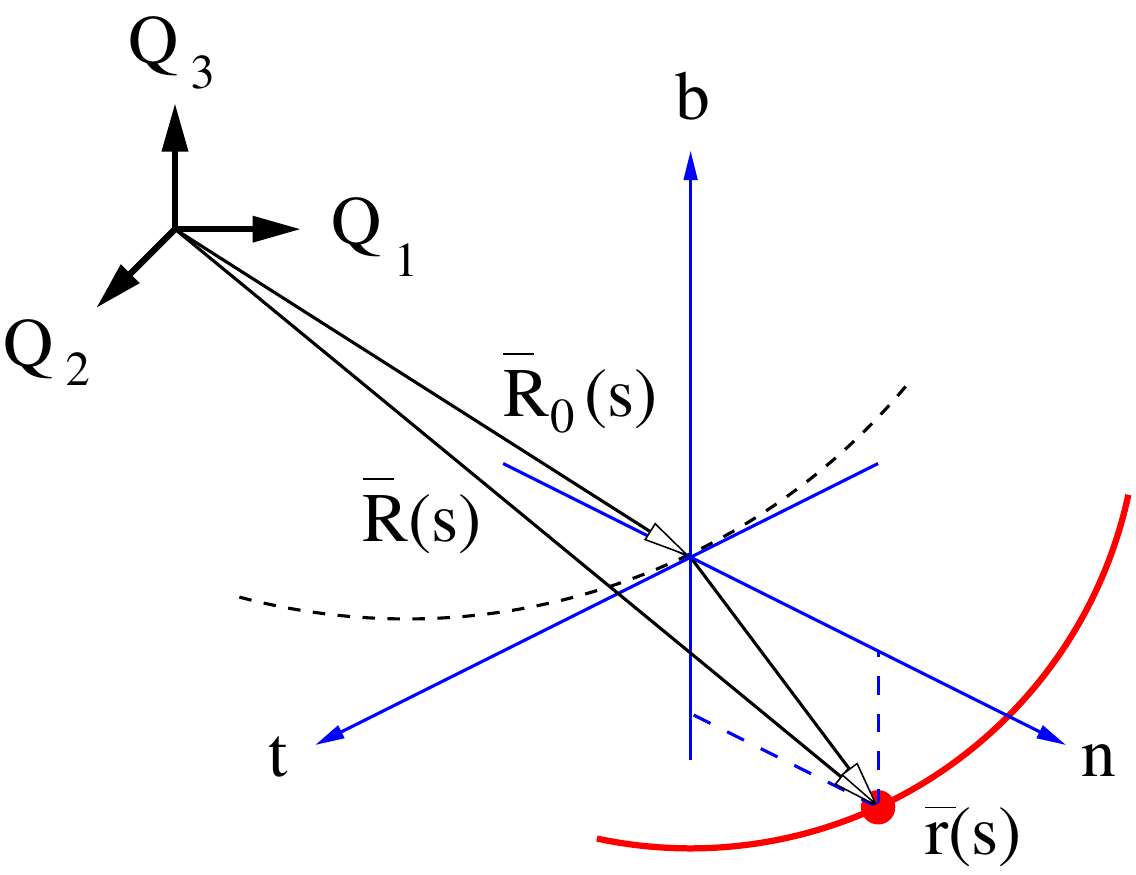}
\caption{\label{fig:SFFrame}
	Illustration of a test particle's position vector expressed as a
	transverse, i.e. for fixed $q_3$, displacement from equilibrium orbit.
	}
\end{figure}

%------------------------------------------------------------------------------%
The {\it equilibrium particle} is a particle with design energy perfectly
following the reference orbit.
Let $\mathbf{R}_0(t)$ be the position vector of it as a function of time.
Then one can describe the equilibrium orbit in terms of its natural
parametrization by arc length as
\[
	s(t) = \int_0^t |\dot{\mathbf{R}}_0(t)|\,\dd\,t.
\]
Now on can introduce the local right-handed orthonormal Frenet-Serret basis
$\{ \nV , \bV , \tV \}$ (or TNB frame), where basis vectors are defined as
follows:
%------------------------------------------------------------------------------%
\begin{itemize}
%------------------------------------------------------------------------------%
\item tangent unit vector
\[
\tV = \frac{\dd\,\mathbf{R}_0(s)}{\dd s},
\]
%------------------------------------------------------------------------------%
\item outward-pointing normal unit vector
\[
\nV = -\frac{1}{\kappa(s)}\frac{\dd\,\tV}{\dd s},
\]
%------------------------------------------------------------------------------%
\item and binormal unit vector
\[
\bV = \tV \times \nV,
\]
\end{itemize}
%------------------------------------------------------------------------------%
where $\kappa = \left|\dd\,\tV/\dd s\right|$ defines the local curvature of
the equilibrium orbit.
Then, using the Frenet-Serret formulas describing the derivatives of unit
vectors in terms of each other
\[
\dd
\begin{bmatrix}
	\tV \\ \nV \\ \bV
\end{bmatrix}
=
\begin{bmatrix}
	 0		& -\kappa	& 0		\\
	 \kappa		& 0		& \tau		\\
	 0		& -\tau		& 0
\end{bmatrix}
\begin{bmatrix}
	\tV \\ \nV \\ \bV\end{bmatrix}
\dd s
\]
where $\tau(s)$ is the torsion of an equilibrium orbit which measures the
failure of a curve to be planar, one can express the position vector of a test
particle as a transverse displacement from equilibrium orbit, 
see FIG.~\ref{fig:SFFrame},
\[
	\mathbf{R}(s)   = \mathbf{R}_0(s) + \mathbf{r}(s) =
	\mathbf{R}_0(s) + q_1\nV + q_2\bV.
\]
and its' infinitesimally small displacement
\[
\dd\mathbf{R} =
	\nV\,\dd q_1 + \bV\,\dd q_2 + (1+\kappa\,q_1)\tV\,\dd q_3 +
	\tau(q_1 \bV - q_2 \nV)\dd q_3,
\]
where $(q_1,q_2,q_3)$ are local curvilinear coordinates spanned on 
$(\nV,\bV,\tV)$.
One can see that in the case of flat orbit, i.e. $\tau = 0$, the local 
Frenet-Serret frame can be associated with global orthogonal coordinate system
with a line element in a form
\[
	\dd\mathbf{l} = \sum_{i=1}^3 h_i \hat{\mathbf{e}}_i \dd q_i,
\]
where scale factors are $h_1 = h_2 = 1$ and $h \equiv h_3 = 1 + \kappa\,q_1$.

%------------------------------------------------------------------------------%
%------------------------------------------------------------------------------%

%------------------------------------------------------------------------------%
The use of global coordinates with metric provided by local Frenet-Serret frame
allows to rewrite the Lagrangian as
\[
\mathcal{L}[\mathbf{r},\dot{\mathbf{r}};t] =
	-m\,c^2\sqrt{1-\frac{v^2}{c^2}} -
	e\,\Phi +
	e\,\mathbf{v}\cdot\mathbf{A},
\]
where $\mathbf{v}=(\dot{q}_1,\dot{q}_2,h\,\dot{q}_3)$ is the particle's velocity
expressed in new coordinates.
Thus the new equations of motion are
\[
	\frac{\dd}{\dd t} (\gamma\,m\,\mathbf{v}) =
	e\left(
		\mathbf{E} + \epsilon_{ijk}\,\hat{\mathbf{e}}_i\,v_j\,B_k
	\right) +
	\gamma\,m\,\dot{q}_3^2\,\mathbf{K}
\]
where the vector in the RHS of equation defined as
\[
	\mathbf{K} = (\kappa\,h,0,\kappa'\,q_1),
\]
and the operator $(\ldots)'\equiv\frac{\dd}{\dd q_3}$ is the derivative with
respect to longitudinal coordinate.
Derivatives of potentials expressed via electromagnetic fields using 
expressions for differential operators in curvilinear orthogonal coordinates 
form Table~\ref{tab:NABLA}.
%------------------------------------------------------------------------------%
Calculating components of the new canonical momenta
\[
\frac{p_i}{h_i} \equiv 
	\frac{1}{h_i} \frac{\pd\,\mathcal{L}}{\pd\dot{q}_i} =
	\gamma\,m\,v_i + e\,\mathbf{A}_i(\mathbf{r})
\]
allows to write down the new Hamiltonian
\[
\mathcal{H}[\mathbf{p},\mathbf{q};t] =
	c\,
	\sqrt{
		m^2c^2 +
		\sum_{i=1}^3
		\left(
			\frac{p_i - e\,h_i A_i}{h_i}
		\right)^2
	} + e\,\Phi
\]
and equations of motion
\begin{eqnarray*}
\dot{q}_i\times h_i	& = &
	\frac{c^2}{\mathcal{H}-e\,\Phi}\,\frac{p_i-e\,h_i A_i}{h_i},     \\
\dot{p}_i\,\,/\,h_i	& = &
	\frac{c^2}{\mathcal{H}-e\,\Phi}
	\left[
		e\,\epsilon_{ijk} \frac{p_j}{h_j}\,B_k +
		\frac{K_i}{h^2}
		\left(\frac{p_3-e\,h\,A_3}{h}\right)^2
	\right]								\\ 
			&   &
	+ e\,E_i.
\end{eqnarray*}

%-------------------------------------------------------------------------------
\begin{table*}[th!]
%-------------------------------------------------------------------------------
\caption{\label{tab:NABLA}Differential operators in general orthogonal
coordinates $(q_1,q_2,q_3)$ where $H = h_1 h_2 h_3$, and its expressions in
orthogonal coordinates associated with Serret-Frenet frame.}
\begin{ruledtabular}
\begin{tabular}{lll}
								\\[-0.4cm]
%-------------------------------------------------------------------------------
Gradient	& $\nabla\phi$			&
$\displaystyle\sum_{k=1}^3
	\frac{1}{h_k}
	\frac{\pd\,\phi}{\pd q^k} \hat{\mathbf{e}}_k$ 		\\[0.4cm]
& & $\displaystyle
	           \frac{\pd\,\phi}{\pd q_1}\,\ex +
	           \frac{\pd\,\phi}{\pd q_2}\,\ey +
	\frac{1}{h}\frac{\pd\,\phi}{\pd q_3}\,\ez$		\\[0.3cm]
\hline								\\[-0.3cm]
%-------------------------------------------------------------------------------
Divergence	& $\nabla\cdot \mathbf{F}$	&
$\displaystyle\sum_{k=1}^3
	\frac{1}{H}\frac{\pd}{\pd q^k}
	\left(
		\frac{H}{h_k} F_k
	\right)$						\\[0.4cm]
& & $\displaystyle\frac{1}{h}
\left[
	\frac{\pd(h\,F_1)}{\pd q_1} +
	\frac{\pd(h\,F_2)}{\pd q_2} +
	\frac{\pd  \,F_3 }{\pd q_3}	
\right]$							\\[0.3cm]
\hline								\\[-0.3cm]
%-------------------------------------------------------------------------------
Curl		& $\nabla\times\mathbf{F}$	&
$\displaystyle\sum_{k=1}^3
	\frac{h_k\,\hat{\mathbf{e}}_k}{H} \epsilon_{ijk}
	\frac{\pd}{\pd q^i}\left( h_j F_j \right)$		\\[0.4cm]
& &$\displaystyle
\frac{1}{h}\left[
	\frac{\pd(h\,F_3)}{\pd q_2} -
	\frac{\pd  \,F_2 }{\pd q_3}
\right]\ex+ 
\frac{1}{h}\left[
	\frac{\pd  \,F_1 }{\pd q_3} -
	\frac{\pd(h\,F_3)}{\pd q_1}
\right]\ey+
\left[
	\frac{\pd\,F_2}{   \pd q_1} -
	\frac{\pd\,F_1}{   \pd q_2}
\right]\ez$							
\\[0.3cm]
\hline								\\[-0.3cm]
%-------------------------------------------------------------------------------
Scalar Laplacian& $\triangle\phi = \nabla \cdot (\nabla\phi)$	&
$\displaystyle \sum_{k=1}^3
	\frac{1}{H}
	\frac{\pd}{\pd q^k}
	\left(
	\frac{H}{h_k^2}\frac{\pd\,\phi}{\pd q^k}
	\right)$						\\[0.4cm]
& & $\displaystyle\frac{1}{h}
\left[
	\frac{\pd}{\pd q_1}	\left(
		h \frac{\pd\,\phi}{\pd q_1}
					\right) +
	\frac{\pd}{\pd q_2}	\left(
		h \frac{\pd\,\phi}{\pd q_2}
					\right) +
	\frac{\pd}{\pd q_3}	\left(
		\frac{1}{h}\frac{\pd\,\phi}{\pd q_3}
					\right)
\right]$							\\[0.3cm]
\hline								\\[-0.2cm]
%-------------------------------------------------------------------------------
Vector Laplacian& $\davidsstar\mathbf{F} =
			\nabla(\nabla \cdot \mathbf{F}) - 
			\nabla \times (\nabla\times\mathbf{F})$	
&
$\displaystyle\sum_{k=1}^3
	\left\{
	\frac{1}{h_k}\frac{\pd}{\pd q^k}
	\left[
		\frac{1}{H}\frac{\pd}{\pd q^i}
		\left(
			\frac{H}{h_i}F_i
		\right)
	\right]
	-
	\frac{h_k}{H}\epsilon_{ijk}\frac{\pd}{\pd q^i}
	\left[
		\frac{h_j^2 \hat{\mathbf{e}}_j}{H} \epsilon_{lmj}
		\frac{\pd}{\pd q^l}(h_m F_m)
	\right]
	\right\}\hat{\mathbf{e}}_k$						
\\[0.1cm]
\end{tabular}
\end{ruledtabular}
\end{table*}
%-------------------------------------------------------------------------------

%==============================================================================%
%==============================================================================%
%==============================================================================%
\section{\label{sec:TEM}Transverse electromagnetic fields}

%------------------------------------------------------------------------------%
Now we will restrict ourself with the case of transverse electromagnetic fields;
in orthogonal curvilinear coordinate system associated with Serret-Frenet frame 
these are the fields with translation symmetry along longitudinal coordinate 
$q_3$.
Thus, the scalar and vector potentials are function of transverse coordinates 
only and vector potential has only one nonvanishing component which is $A_3$.
Both potentials satisfies Laplace equation
\begin{eqnarray*}
\triangle \Phi		& =    &
\frac{1}{h}
\left[
	\frac{\pd}{\pd q_1}	\left(
		h \frac{\pd\,\Phi}{\pd q_1}
					\right) +
	\frac{\pd}{\pd q_2}	\left(
		h \frac{\pd\,\Phi}{\pd q_2}
					\right)
\right]=0,								\\
\davidsstar \mathbf{A}	& =    &
	\frac{\pd}{\pd q_1}	\left[\frac{1}{h}
		\frac{\pd\,(h\,A_3)}{\pd q_1}
					\right] +
	\frac{\pd}{\pd q_2}	\left[\frac{1}{h}
		\frac{\pd\,(h\,A_3)}{\pd q_2}
					\right] =0.
\end{eqnarray*}
The corresponding fields are given by Maxwell equations
\[
\mathbf{E} = -\nabla \Phi
\qquad\text{and}\qquad
\mathbf{B} =\nabla \times \mathbf{A}
\]
with differential operators defined for orthogonal curvilinear coordinate
system (Table~\ref{tab:NABLA}), and one gets
\begin{eqnarray*}
E_1 &=& -\frac{\pd\,\Phi}{\pd q_1},
	\qquad\qquad
B_1 = \,\,\,\,\frac{1}{h}\frac{\pd(h\,A_3)}{\pd q_2},		\\
E_2 &=& -\frac{\pd\,\Phi}{\pd q_2},
	\qquad\qquad
B_2 = -\frac{1}{h}\frac{\pd(h\,A_3)}{\pd q_1}.
\end{eqnarray*}

%==============================================================================%
%==============================================================================%
%==============================================================================%
\subsection{\label{sec:t-R}$t$-representation}

%------------------------------------------------------------------------------%
In the case of pure electric or magnetic fields further simplifications can be
applied.
For numerical integration purposes it is very convenient to have a Hamiltonian
in a form of a sum of ``kinetic'' and ``potential'' energies where potentials
will be separated from momentum variables.
In this case, one can easily construct symplectic integrator consisting of
``drifts'' and ``kicks'' associated with kinetic and potential terms 
respectively (e.g.~\cite{Yoshida:1990zz}).

%------------------------------------------------------------------------------%
For pure electric field when curvature is independent of longitudinal coordinate
not only Hamiltonian but also $p_3$ is an invariant of motion, and, problem is 
essentially two dimensional.
Measuring the time in units of $c\,t$ and normalizing the transverse momentums
over the longitudinal component, $\tilde{p}_{1,2} = p_{1,2}/p_3$, one has
\[
\mathrm{H}[\tilde{\mathbf{p}},\mathbf{q};c\,t] =
	\frac{1}{h}\sqrt{
		\frac{p_3^2+h^2m^2c^2}{p_3^2} +
		h^2(\tilde{p}_1^2 + \tilde{p}_2^2)
	} + \frac{e}{p_3\,c}\,\Phi.
\]
We will call this model Hamiltonian the {\it $t$-representation};
with no assumptions made, but the field symmetry, we derived general equations
of motion which can be used for the basis for the construction of symplectic 
integrator.
In a paraxial approximation, $\tilde{p}_{1,2}\ll 1$,and for
$p_{1,2} \gg m\,c$ 
the form is significantly simpler, and a limit of straight coordinates when 
$h=1$ is obvious
\[
\mathrm{H}[\tilde{\mathbf{p}},\mathbf{q};c\,t] \approx
	h\,
	\left(
		\frac{\tilde{p}_1^2}{2} + \frac{\tilde{p}_2^2}{2}
	\right)
	+\frac{1}{h}
	+ \frac{e}{p_3\,c}\,\Phi.
\]

%==============================================================================%
%==============================================================================%
%==============================================================================%
\subsection{\label{sec:s-R}$s$-representation}

%------------------------------------------------------------------------------%
For pure magnetic field the Hamiltonian is very hard to exploit since it has 
only a square root and so no terms to split.
Introducing an extended Hamiltonian with a new fictitious time parameter, 
$\tau$, where the old independent variable and old Hamiltonian with a negative 
sign will be treated as an additional pair of canonically conjugated
coordinates, $(-\mathcal{H},t)$, one have:
\begin{eqnarray*}
0	&\equiv&
	\mathcal{O} [p_1,p_2,p_3,-\mathcal{H};q_1,q_2,q_3,t;\tau]	\\
	& =    &
	c\,\sqrt{
		m^2c^2 + p_1^2 + p_2^2 +
	      \left(
	      \frac{p_3 - e\,h\,A_3}{h}
	\right)^2 } -
	\mathcal{H}.
\end{eqnarray*}
Integration of additional equations of motion gives
\[
\mathcal{H} = \mathrm{inv}
\qquad\text{and}\qquad
t	    = \tau + C_0,
\]
where we can set a constant of integration $C_0=0$.

%------------------------------------------------------------------------------%
If curvature is invariant of longitudinal coordinate the longitudinal component
of momentum conserved, as well as in the case of electric field, and we will
use $-p_3$ as a new Hamiltonian, reducing number of degrees of freedom back up 
to three by using $q_3$ as a new independent variable:
\begin{eqnarray*}
-p_3	&\equiv&
	\mathcal{K}[p_1,p_2,-\mathcal{H};q_1,q_2,t;q_3]		\\
	&=&    
	- h\,\sqrt{
		\left(\frac{\mathcal{H}}{c}\right)^2 - m^2c^2 -
		p_1^2 - p_2^2
	}          - e\,h\,A_3.
\end{eqnarray*}
The use of generating function
\[
	G_2(t,-\Pi) = - t \sqrt{\Pi^2c^2+(m\,c^2)^2}
\]
will allow to use the full kinetic momentum $-\Pi$ of a particle instead of 
$-\mathcal{H}$ as one of canonical momentums:
\[
	\mathcal{K}[p_1,p_2,-\Pi;q_1,q_2,l;q_3] =
	- h\,\sqrt{ \Pi^2 - p_1^2 - p_2^2}
	- e\,h\,A_3,
\]
where corresponding canonical coordinate is a particle's traversed path
$l = -\pd\,G_2/\pd\Pi = \beta c\,t.$
%\[
%	l = \frac{\partial\,G_2(t,-\Pi)}{\partial(-\Pi)} = \beta c\,t.
%\]

%------------------------------------------------------------------------------%
Since the Hamiltonian do not explicitly depends on $l$, full momentum $\Pi$ is
conserved and we can exclude associated degree of freedom using the further
renormalization of the Hamiltonian
$\mathcal{K} \rightarrow \mathrm{K} \equiv \mathcal{K}/\Pi$, which can be 
achieved by re-normalizing transverse components of canonical momentums
$p_{1,2} \rightarrow \tilde{p}_{1,2} = p_{1,2}/\Pi$:
\begin{eqnarray*}
-\frac{p_3}{\Pi}	&\equiv&
	\mathrm{K}[\tilde{p}_1,\tilde{p}_2;q_1,q_2;q_3]			\\
			& = &
	-h\,\sqrt{
			1 - \tilde{p}_1^2 - \tilde{p}_2^2
	} - \frac{e}{\Pi}\,h\,A_3.
\end{eqnarray*}
We will call this model Hamiltonian {\it $s$-representation} since the 
longitudinal coordinate (sometimes referred to the natural parameter along 
equilibrium orbit, $s$) is used as a time-parameter.
This representation is convenient to use for the numerical integrator 
construction for transverse magnetic fields.
The paraxial approximation, $\tilde{p}_{1,2}\ll 1$, gives
\[
\mathrm{K}[\tilde{\mathbf{p}},\mathbf{q};q_3] \approx
	h\,
	\left(
		\frac{\tilde{p}_1^2}{2} + \frac{\tilde{p}_2^2}{2}
	\right)
	-h
	- \frac{e}{\Pi}\,h\,A_3.
\]

%==============================================================================%
%==============================================================================%
%==============================================================================%
\subsection{R- and S-elements}

%------------------------------------------------------------------------------%
So far we provided dynamical equations of motion without specifying how to
represent electromagnetic fields.
In next two subsections we will discuss the multipole field expansion for two 
most important types of elements: R-element for $\kappa=0$ and S-element defined
for $\kappa = \text{const} = 1/R_0$.

%------------------------------------------------------------------------------%
R- stays for rectangular and this element is the one whit $(q_1,q_2,q_3)$ 
simply being the right handed Cartesian coordinate system which we will denote 
as $(x,y,z)$.
All fields in such an element are invariant along $z$ axis and usually serves 
the function of regular quadrupoles, sextupoles, octupoles or combined function 
correctors.
In addition one can design pure R-dipoles, while combined function bending 
magnets are exotic and very complicated since equilibrium orbit will not anymore
coincides with axis of symmetry.

\begin{figure}[b!]
\includegraphics[width=0.87\linewidth]{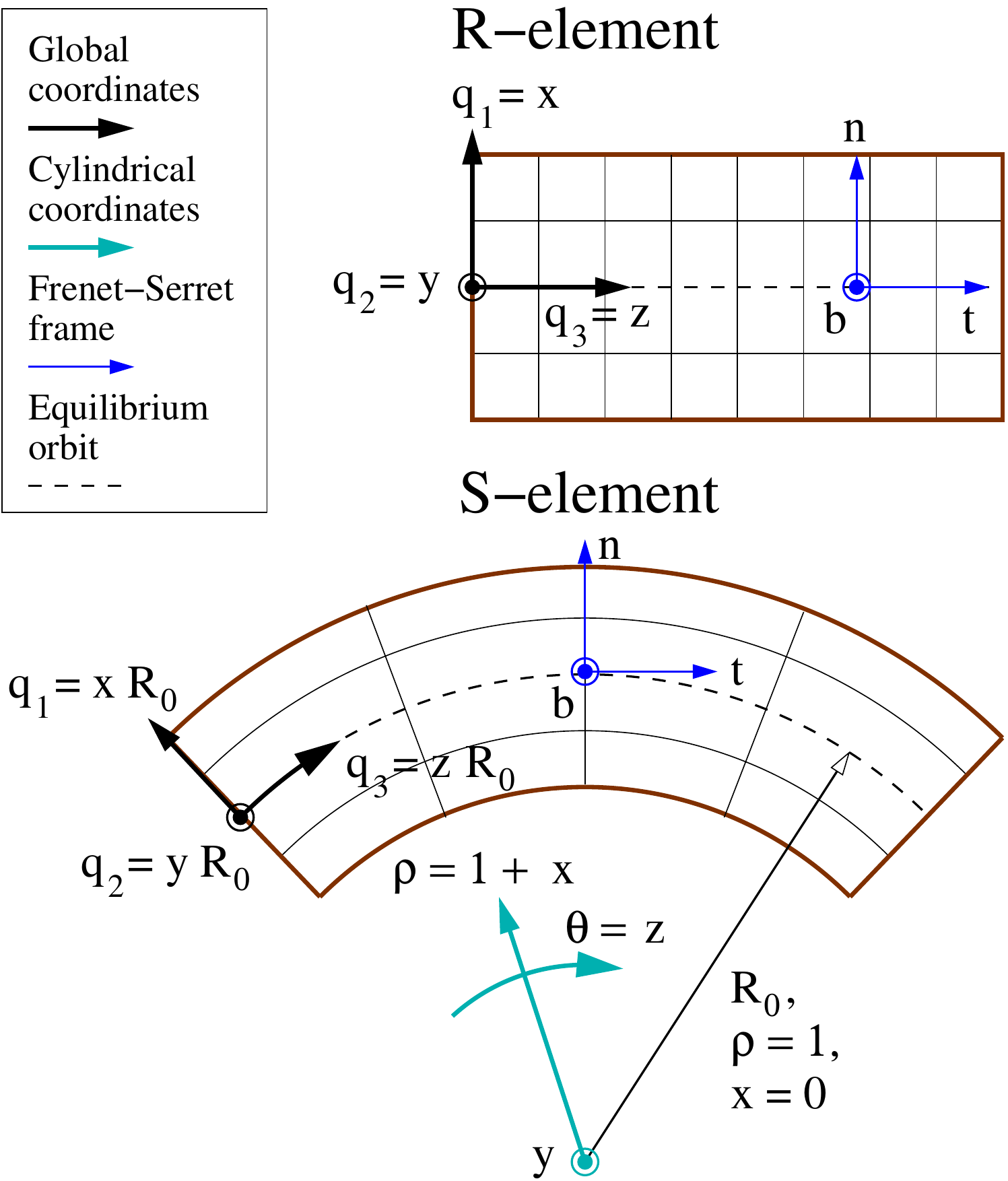}
\caption{\label{fig:RSElem} Illustration of R- and S- elements.
	Elements are shown in brown.
	Global curvilinear coordinates with associated grid lines are shown in
	black.
	Black dashed line represent an equilibrium orbit.
	An example of Frenet-Serret frame attached to an equilibrium orbit
	drawn in blue colors.
	For S-element, an additional right-handed normalized cylindrical system
	is added and shown in cyan.}
\end{figure}

%------------------------------------------------------------------------------%
S-element is the element defined whit natural sector coordinate system.
Defining the set of normalized coordinates $(x=q_1/R_0,y=q_2/R_0,z=q_3/R_0)$,
one can see that it simply can be related to normalized right handed cylindrical
coordinates $(\rho = 1+x,y,\theta=z/R_0)$, see FIG.~\ref{fig:RSElem},
and thus all fields are invariant along azimuthal coordinate $\theta$.
S-elements are suitable for the design of combined function bending magnets,
since in contrast to R-elements, equilibrium orbit follows along $\theta$.

%------------------------------------------------------------------------------%
%------------------------------------------------------------------------------%

\begin{table*}[t!]
\caption{\label{tab:AB}Harmonic homogeneous polynomials in two variables.}
\begin{ruledtabular}
\begin{tabular}{lll}
$n$	& $\mathcal{A}_n$		& $\mathcal{B}_n$	\\\hline
0	& 1
	& 0							\\
1	& $x$
	& $y$							\\
2	& $x^2-y^2$
	& $2\,x\,y$						\\
3	& $x^3-3\,x\,y^2$
	& $3\,x^2y-y^3$						\\
4	& $x^4-6\,x^2y^2+y^4$
	& $4\,x^3y-4\,x\,y^3$					\\
5	& $x^5-10\,x^3y^2+5\,x\,y^4$
	& $5\,x^4y-10\,x^2y^3+y^5$				\\
6	& $x^6-15\,x^4y^2+15\,x^2y^4-y^6$
	& $6\,x^5y-20\,x^3y^3+6\,x\,y^5$			\\
7	& $x^7-21\,x^5y^2+35\,x^3y^4-7\,x\,y^6$
	& $7\,x^6y-35\,x^4y^3+21\,x^2y^5-y^7$			\\
8	& $x^8-28\,x^6y^2+70\,x^5y^4-84\,x^3y^6+9\,x\,y^8$
	& $8\,x^7y-56\,x^5y^3+56\,x^3y^5-8\,x\,y^7$		\\
9	& $x^9-36\,x^7y^2+126\,x^5y^4-84\,x^3y^6+9\,x\,y^8$
	& $9\,x^8y-84\,x^6y^3+126\,x^4y^5-36\,x^2y^7+y^9$	\\
\end{tabular}
\end{ruledtabular}
\end{table*}

%==============================================================================%
%==============================================================================%
\subsection{\label{sec:RM}Multipoles in Cartesian coordinates}

%------------------------------------------------------------------------------%
In Cartesian coordinates Laplace equations for electro- and magnetostatic fields
are in the same form which significantly simplify the problem
\begin{eqnarray*}
\triangle_{\perp}   \Phi	&=&
	\frac{\pd^2 \Phi}{\pd x^2}+ \frac{\pd^2 \Phi}{\pd y^2}=0,\\
\davidsstar_{\perp} \mathbf{A}	&=&
	\left(
	\frac{\pd^2 A_z}{\pd x^2} + \frac{\pd^2 A_z}{\pd y^2}
	\right)\hat{\mathbf{e}}_z =0.
\end{eqnarray*}
Introduction of complex variables allows a very compact description of a 
problem with unified description of electric and magnetic fields.
Suppose we have a holomorphic function of complex variable $\mathcal{Z}=x+i\,y$
which we will call {\it complex scalar potential} which real part is defined to
be a longitudinal component of a vector potential and imaginary part is the 
electric scalar potential
\[
	\Omega(\mathcal{Z}) = A_z(x,y) + i\,\Phi(x,y).
\]
Since real or imaginary part of any holomorphic function are harmonic functions,
$A_z$ and $\Phi$ automatically satisfies the Laplace equation.
Indeed, suppose we have a vector field $\mathbf{F}=(F_x,F_y)$.
Introducing the Wirtinger derivatives
\[
\frac{\pd}{\pd \mathcal{Z}} = \frac{1}{2}	\left(
	\frac{\pd}{\pd x} -i\,\frac{\pd}{\pd y}
				\right)
\quad\text{and}\quad
\frac{\pd}{\pd \overline{\mathcal{Z}}} = \frac{1}{2}	\left(
	\frac{\pd}{\pd x} +i\,\frac{\pd}{\pd y}
				\right)
\]
one can write
\begin{eqnarray*}
	\frac{\pd\,\Omega}{\pd \overline{\mathcal{Z}}} &=& 0,			
	\\
	\frac{\pd\,\Omega}{\pd      \mathcal{Z} } &=& F(\mathcal{Z}),
\end{eqnarray*}
where first equation is the Cauchy-Riemann condition for $\Omega$ which
guarantees that this field can be implemented via either magnetic or electric
potentials:
\begin{eqnarray*}
	F_x &=& -\frac{\pd\,\Phi}{\pd x} =\,\,\,\,\frac{\pd\,A_z}{\pd y}, 
\\
	F_y &=& -\frac{\pd\,\Phi}{\pd y} =-       \frac{\pd\,A_z}{\pd x}.
\end{eqnarray*}
The second equation defines complex function of field components such that
\[
F_x = -\Im\,F(\mathcal{Z})
\quad\text{and}\quad
F_y = -\Re\,F(\mathcal{Z}),
\]
which all together are equivalent to
$\mathbf{F} = -\nabla \Phi = \nabla\times\mathbf{A}$.
The complex function $F(\mathcal{Z})$ is the holomorphic function again and
Cauchy-Riemann equation gives
\[
	\frac{\pd\,F}{\pd \overline{\mathcal{Z}}} = 0,
\]
that asserts that field $\mathbf{F}$ is irrotational and divergence free 
which is equivalent to time-independent free of electric charge and current 
densities Maxwell's equations
\[
\nabla \cdot  \mathbf{F} = 0
\quad\text{and}\quad
\nabla \times \mathbf{F} = 0.
\]

%------------------------------------------------------------------------------%
\begin{figure*}[th!]
\includegraphics[width=\linewidth]{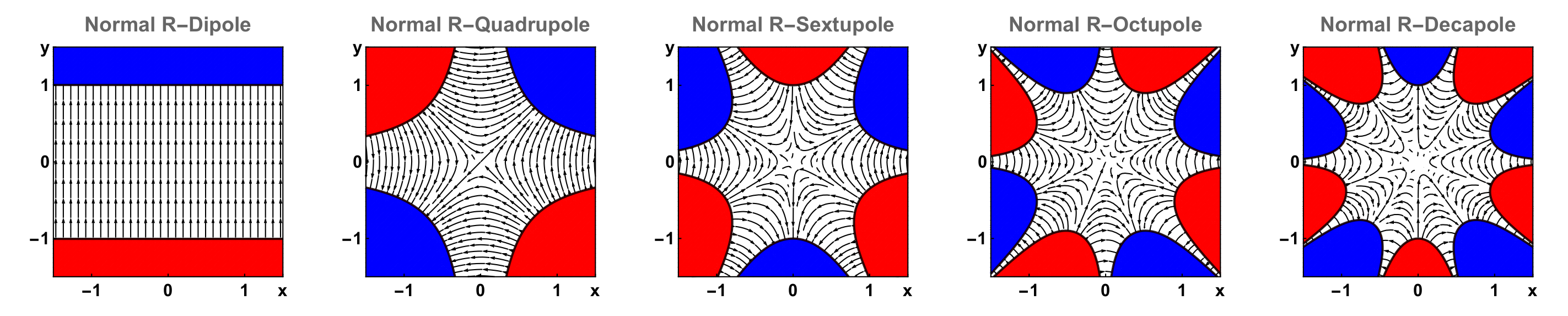}
\includegraphics[width=\linewidth]{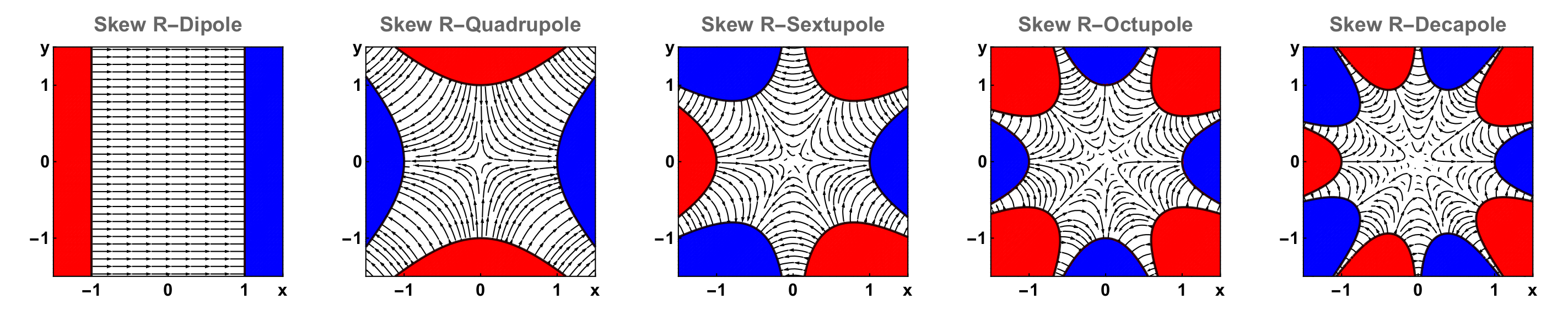}
\caption{\label{fig:RMult} Normal and skew $2n$-pole magnets in Cartesian
	coordinates.
	Each figure shows magnetic (electric) field streamlines and poles' shape
	in transverse cross section.
	North (positive electrostatic potential) and
	south (negative electrostatic potential) poles are shown in red and blue
	and are given by $(\mathcal{B,A})_n = \mp R_{\text{p}}^n$ respectively, 
	where $R_{\text{p}}$ is the distance to the pole's tip.}
\end{figure*}
%------------------------------------------------------------------------------%

%------------------------------------------------------------------------------%
For accelerator physics purposes the expansion of fields usually represented
in terms of homogeneous harmonic polynomials of two variables, which are defined
through the complex power function
\begin{eqnarray*}
	\An (x,y)	&=& \Re\,\mathcal{Z}^n =
	\frac{1}{2}
	\left[
		\left(x + i\,y\right)^n + \left(x - i\,y\right)^n
	\right]								\\
		&=& \sum_{k=0}^n \begin{pmatrix} n \\ k \end{pmatrix}
	x^{n-k}y^k\cos\frac{k\,\pi}{2},					\\
	\Bn (x,y)	&=& \Im\,\mathcal{Z}^n =
	\frac{1}{2\,i}
	\left[
		\left(x + i\,y\right)^n - \left(x - i\,y\right)^n
	\right]								\\
		&=& \sum_{k=0}^n \begin{pmatrix} n \\ k \end{pmatrix}
	x^{n-k}y^k\sin\frac{k\,\pi}{2}.
\end{eqnarray*}
Explicit expressions up to 10-th order are in Table~\ref{tab:AB}.
These functions satisfy the Laplace equation
$\triangle_\perp=0$
and related to each other through Cauchy-Riemann equation as
\[
\frac{\pd\,\An}{\pd x} = \frac{\pd\,\Bn}{\pd y}
\quad\text{and}\quad
\frac{\pd\,\An}{\pd y} =-\frac{\pd\,\Bn}{\pd x}.
\]
In addition one can introduce ``ladder-like'' lowering differential operators as
\[
n\,\left\{\mathcal{A,B}\right\}_{n-1} =
	   \frac{\pd}{\pd x}\left\{\mathcal{A,B}\right\}_n =
	\pm\frac{\pd}{\pd y}\left\{\mathcal{B,A}\right\}_n.
\]
%\begin{eqnarray*}
%\An &=& \frac{1}{n+1}\frac{\pd}{\pd x}\mathcal{A}_{n+1}
%		=\,\,\,\,\frac{1}{n+1}\frac{\pd}{\pd y}\mathcal{B}_{n+1},\\
%\Bn &=& \frac{1}{n+1}\frac{\pd}{\pd x}\mathcal{B}_{n+1}
%		=-\frac{1}{n+1}\frac{\pd}{\pd y}\mathcal{A}_{n+1}.
%\end{eqnarray*}
%and
%\begin{eqnarray*}
%\An &=&	n\,\int_{0}^{x}\dd x\,\mathcal{A}_{n-1} +
%			y^n\cos\frac{n\,\pi}{2}
%		=	x^n-n\,\int_{0}^{y}\dd y\,\mathcal{B}_{n-1}	\\
%\Bn &=&	n\,\int_{0}^{x}\dd x\,\mathcal{B}_{n-1}\,+
%			y^n\sin\frac{n\,\pi}{2}
%		=	n\,\int_{0}^{y}\dd y\,\mathcal{A}_{n-1}.	\\
%			\end{eqnarray*}
%with $\mathcal{A}_0=1$ and $\mathcal{B}_0=0$.

%------------------------------------------------------------------------------%
Thus one can define two independent of each other sets of solutions,
{\it normal} (sometimes called {\it upright} or {\it straight}) and {\it skew}
pure multipoles, which we will denote with overline $\overline{(\ldots)}$ and
underline $\underline{(\ldots)}$ respectively.
The complex scalar potentials of pure multipoles are:
\[
	\overline{\Omega} ^{(n)} = -\,\,\Cno\frac{\mathcal{Z}^n}{n!}
	\qquad\text{and}\qquad
	\underline{\Omega}^{(n)} = -i\, \Cnu\frac{\mathcal{Z}^n}{n!}
\]
where $\Cno$ and $\Cnu$ are coefficients determining the strength of magnets.
Corresponding vector fields are defined to have an odd and even {\it midplane
symmetries}
\begin{eqnarray*}
	\Fyon (x,y) &=& \Fyon (x,-y)
	\qquad\text{and}\qquad
	\Fxon (x,0) = 0,				\\
	\Fxun (x,y) &=& \Fxun (x,-y)
	\qquad\text{and}\qquad
	\Fyun (x,0) = 0.
\end{eqnarray*}
Formulas for potentials and fields are listed below in Table~\ref{tab:R-El} and
exact expressions are provided in Appendix~\ref{secAP:NSMult}.
Figure~\ref{fig:RMult} shows the cross section of idealized multipole magnet's 
poles and corresponding fields.

%------------------------------------------------------------------------------%
\begin{table}[h!]
\caption{\label{tab:R-El}
	Formulas for the scalar potential, longitudinal component of the vector 
	potential and field components for pure normal and skew $2n$-poles in
	Cartesian coordinates.}
\begin{ruledtabular}
\begin{tabular}{p{0.45\columnwidth}|p{0.5\columnwidth}}
Normal	& Skew								\\\hline
									\\
[-0.3cm]
$\displaystyle\FIon = -  \Cno\frac{\Bn}{n!}$				&
$\displaystyle\FIun = -  \Cnu\frac{\An}{n!}$				\\
[0.2cm]
$\displaystyle\Ason = -  \Cno\frac{\An}{n!}$				&
$\displaystyle\Asun =\quad\!\!\Cnu\frac{\Bn}{n!}$			\\
[-0.3cm]
									\\\hline
									\\
[-0.3cm]
$\displaystyle\Fxon =\quad\!\!\Cno\frac{\mathcal{B}_{n-1}}{(n-1)!}$	&
$\displaystyle\Fxun =\quad\!\!\Cnu\frac{\mathcal{A}_{n-1}}{(n-1)!}$	\\
[0.2cm]
$\displaystyle\Fyon =\quad\!\!\Cno\frac{\mathcal{A}_{n-1}}{(n-1)!}$	&
$\displaystyle\Fyun = -   \Cnu\frac{\mathcal{B}_{n-1}}{(n-1)!}$
%------------------------------------------------------------------------------%
\end{tabular}
\end{ruledtabular}
\end{table}
%------------------------------------------------------------------------------%

%------------------------------------------------------------------------------%
Therefore, if one provided with experimental data of the power series expansions
of the fields in a horizontal or vertical planes
\begin{eqnarray*}
\left. F_x \right|_{x=0} &=&
	\left.              F_x           \right|_{\text{eq}} +
	\frac{y  }{1!}\,\left. \frac{\pd  \,F_x}{\pd y  } \right|_{\text{eq}} +
	\frac{y^2}{2!}\,\left. \frac{\pd^2\,F_x}{\pd y^2} \right|_{\text{eq}} +
	\ldots,						\\
\left. F_y \right|_{x=0} &=&
	\left.              F_y           \right|_{\text{eq}} +
	\frac{y  }{1!}\,\left. \frac{\pd  \,F_y}{\pd y  } \right|_{\text{eq}} +
	\frac{y^2}{2!}\,\left. \frac{\pd^2\,F_y}{\pd y^2} \right|_{\text{eq}} +
	\ldots,						\\
\left. F_x \right|_{y=0} &=&
	\left.              F_x           \right|_{\text{eq}} +
	\frac{x  }{1!}\,\left. \frac{\pd  \,F_x}{\pd x  } \right|_{\text{eq}} +
	\frac{x^2}{2!}\,\left. \frac{\pd^2\,F_x}{\pd x^2} \right|_{\text{eq}} +
	\ldots,						\\
\left. F_y \right|_{y=0} &=&
	\left.              F_y           \right|_{\text{eq}} +
	\frac{x  }{1!}\,\left. \frac{\pd  \,F_y}{\pd x  } \right|_{\text{eq}} +
	\frac{x^2}{2!}\,\left. \frac{\pd^2\,F_y}{\pd x^2} \right|_{\text{eq}} +
	\ldots,						\\
\end{eqnarray*}
the field derivatives on equilibrium orbit can be related to strength 
coefficients, see Table~\ref{tab:CnR}, which allows to expand a general 
R-element in terms of pure multipoles.

%------------------------------------------------------------------------------%
%\begingroup
%\squeezetable
\begin{table}[b!]
\caption{\label{tab:CnR}
	Relationship between coefficients determining the strength of pure
	R-multipoles and power series expansion of field in horizontal
	and vertical planes on equilibrium orbit.}
\begin{ruledtabular}
\begin{tabular}{lcccc}
%------------------------------------------------------------------------------%
    &\multicolumn{2}{c}{$x=0$}	& \multicolumn{2}{c}{$y=0$}	
\\\hline
$n$ & $\Cno$	& $\Cnu$	& $\Cno$	& $\Cnu$	\\\hline
%------------------------------------------------------------------------------%
1 &
$	F_y$				&
$	F_x$				&
$	F_y$				&
$	F_x$				\\[0.1cm]
2 &
$	\pd_y\,F_x$			&
$-	\pd_y\,F_y$			&
$	\pd_x\,F_y$			&
$	\pd_x\,F_x$			\\[0.1cm]
3 &
$-	\pd_y^2\,F_y$			&
$-	\pd_y^2\,F_x$			&
$	\pd_x^2\,F_y$			&
$	\pd_x^2\,F_x$			\\[0.1cm]
4 &
$-	\pd_y^3\,F_x$			&
$	\pd_y^3\,F_y$			&
$	\pd_x^3\,F_y$			&
$	\pd_x^3\,F_x$			\\[0.1cm]
5 &
$	\pd_y^4\,F_y$			&
$	\pd_y^4\,F_x$			&
$	\pd_x^4\,F_y$			&
$	\pd_x^4\,F_x$			\\[0.1cm]
%6 &
%$	\pd_y^5\,F_x$			&
%$-	\pd_y^5\,F_y$			&
%$	\pd_x^5\,F_y$			&
%$	\pd_x^5\,F_x$			\\[0.1cm]
%7 &
%$-	\pd_y^6\,F_y$			&
%$-	\pd_y^6\,F_x$			&
%$	\pd_x^6\,F_y$			&
%$	\pd_x^6\,F_x$
%8 &
%$-	\pd_y^7\,F_x$			&
%$	\pd_y^7\,F_y$			&
%$	\pd_x^7\,F_y$			&
%$	\pd_x^7\,F_x$			\\[0.1cm]
%9 &
%$	\pd_y^8\,F_y$			&
%$	\pd_y^8\,F_x$			&
%$	\pd_x^8\,F_y$			&
%$	\pd_x^8\,F_x$			\\
%------------------------------------------------------------------------------%
\end{tabular}
\end{ruledtabular}
\end{table}
%\endgroup
%------------------------------------------------------------------------------%

% MULTIPOLES IN SECTOR COORDINATES ============================================%
%==============================================================================%
%==============================================================================%
%==============================================================================%
%==============================================================================%
\subsection{\label{sec:SM}Multipoles in cylindrical coordinates}

%------------------------------------------------------------------------------%
In the normalized right-handed cylindrical coordinate system the Laplace 
equations are
\begin{eqnarray*}
\triangle_{\curvearrowright} \Phi		& = &
	\triangle_\perp\Phi + \frac{1}{\rho}\frac{\pd\,\Phi}{\pd \rho}	\\
	& = &
	\frac{\pd^2\Phi}{\pd \rho^2} +
	\frac{1}{\rho}\frac{\pd\,\Phi}{\pd \rho} +
	\frac{\pd^2\Phi}{\pd y^2}=0,					\\
\davidsstar_{\curvearrowright} \mathbf{A}	& = &
	\left(
		\triangle_{\curvearrowright} \At - \frac{\At}{\rho^2}
	\right) \hat{\mathbf{e}}_\theta					\\
	& = &
	\left(
		\frac{\pd^2 \At}{\pd \rho^2} +
		\frac{1}{\rho}\frac{\pd\,\At}{\pd \rho} +
		\frac{\pd^2 \At}{\pd y^2} -
		\frac{\At}{\rho^2}
	\right) \hat{\mathbf{e}}_\theta = 0.
\end{eqnarray*}
Compared to the case with Cartesian coordinates these equations look quite 
different from each other.
In order to retain the symmetry one can note that
\[
(\davidsstar_{\curvearrowright} \mathbf{A})_\theta = 
	\frac{1}{\rho} \left[
		\frac{\pd^2}{\pd \rho^2} -
		\frac{1}{\rho}\frac{\pd}{\pd \rho} +
		\frac{\pd^2}{\pd    y^2}
	\right] \left(\rho\,\At\right).
\]
Thus looking for the solution in a form similar to harmonic homogeneous 
polynomials
\begin{eqnarray*}
\Phi	&=&
-\sum_{k=0}^n\phantom{\frac{1}{\rho}}
\frac{\mathcal{F}_{n-k}(\rho)}{(n-k)!} \frac{y^k}{k!}	
\left(
	\Cno\,\sin \frac{k\,\pi}{2} +
	\Cnu\,\cos \frac{k\,\pi}{2}
\right),								\\
\At	&=&
-\sum_{k=0}^n
\frac{1}{\rho}\frac{\mathcal{G}_{n-k}(\rho)}{(n-k)!} \frac{y^k}{k!}
\left(
	\Cno\,\cos \frac{k\,\pi}{2} -
	\Cnu\,\sin \frac{k\,\pi}{2}
\right),
\end{eqnarray*}
where $\Fn(\rho)$ and $\Gn(\rho)$ are the functions to be
determined, one can find two recurrence equations
\begin{eqnarray*}
\frac{\pd^2 \Fn(\rho)}{\pd \rho^2} +
	\frac{1}{\rho}\frac{\pd\,\Fn(\rho)}{\pd \rho} &=& 
	n\,(n-1)\,\mathcal{F}_{n-2}(\rho),				\\
\frac{\pd^2 \Gn(\rho)}{\pd \rho^2} -
	\frac{1}{\rho}\frac{\pd\,\Gn(\rho)}{\pd \rho} &=& 
	n\,(n-1)\,\mathcal{G}_{n-2}(\rho).				\\
\end{eqnarray*}
They relate $\Fn$ and $\Gn$ to each other through
\[
\mathcal{G}_{n-1} = \frac{1}{n}\,\rho\,
	\frac{\pd\,\Fn}{\pd \rho}
\qquad\text{and}\qquad
\mathcal{F}_{n-1} = \frac{1}{n}\frac{1}{\rho}
	\frac{\pd\,\Gn}{\pd \rho}.
\]
and allows to construct lowering operators
\begin{eqnarray*}
\Fn &=& \frac{1}{(n+1)(n+2)}
\left[
\frac{1}{\rho}\frac{\pd}{\pd \rho}\left(\rho\,\frac{\pd}{\pd \rho} \right)
\right]
\mathcal{F}_{n+2},							\\
\Gn &=& \frac{1}{(n+1)(n+2)}
\left[
\rho\,\frac{\pd}{\pd \rho}\left(\frac{1}{\rho}\frac{\pd}{\pd \rho} \right)
\right]
\mathcal{G}_{n+2},
\end{eqnarray*}
and thus defines raising operators
\begin{eqnarray*}
\Fn &=& n\,(n-1)
\int_1^\rho \frac{1}{\rho} \int_1^\rho 
\rho\,\mathcal{F}_{n-2}\,\dd\,\rho\,\dd\,\rho,
\\
\Gn &=& n\,(n-1)
\int_1^\rho \rho \int_1^\rho 
\frac{1}{\rho}\,\mathcal{G}_{n-2}\,\dd\,\rho\,\dd\,\rho,
\end{eqnarray*}
where limits of integration are taking care of two constants of integration.
These operators can be used to recursively calculate all members of
$\mathcal{F}-$ and $\mathcal{G}-$functions;
an additional constraint to terminate recurrences defines lowest orders $n=0,1$
as
\[
\mathcal{F}_0 = 1,	\quad \mathcal{F}_1 = \ln\,\rho,	\quad
\mathcal{G}_0 = 1,	\quad \mathcal{G}_{\,1}= (\rho^2-1)/2.
\]
First ten members of $\Fn$ and $\Gn$ are listed in 
Tables~\ref{tab:Fn},~\ref{tab:Gn} and are shown in FIG.~\ref{fig:FG};
in Appendix~\ref{secAP:Taylor} one can find Taylor series of these
functions at $\rho=1$.
The difference relation for $\Fn$ including first members have been found
by~E.M.~McMillan and I would like to acknowledge his result by given them a
name of {\it McMillan radial harmonics}.
In addition to his results, {\it adjoint McMillan radial harmonics}, $\Gn$, are
introduced in order to provide the symmetry in description between electric and
magnetic fields.

\begin{figure*}[p!]
\includegraphics[width=\linewidth]{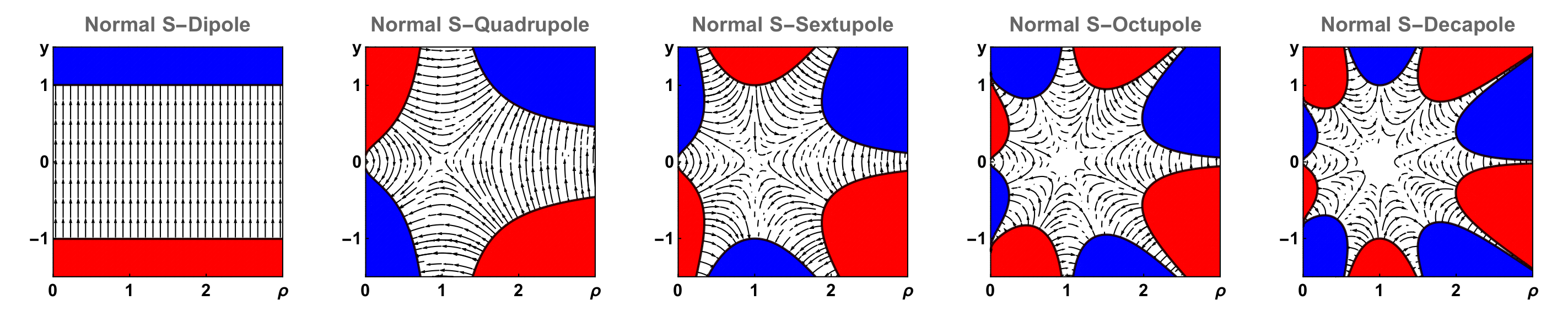}
\includegraphics[width=\linewidth]{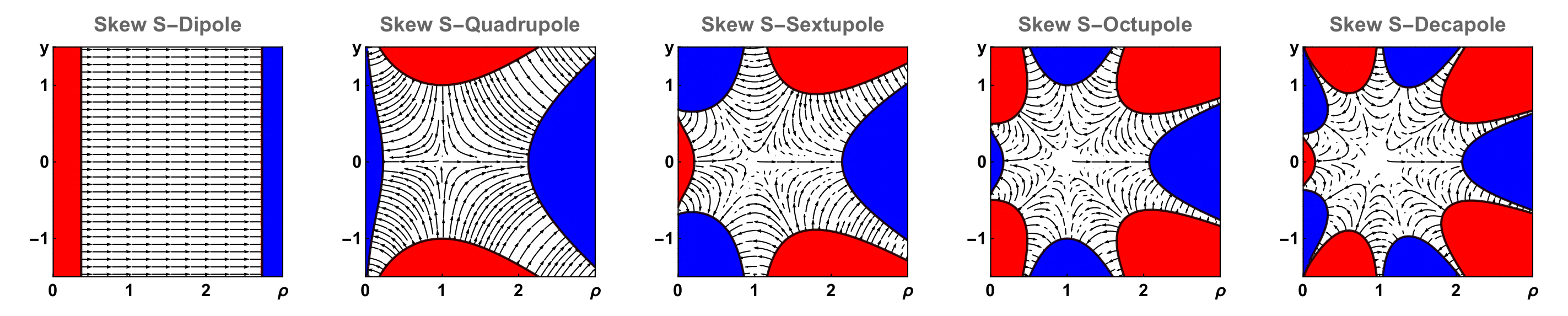}
\includegraphics[width=\linewidth]{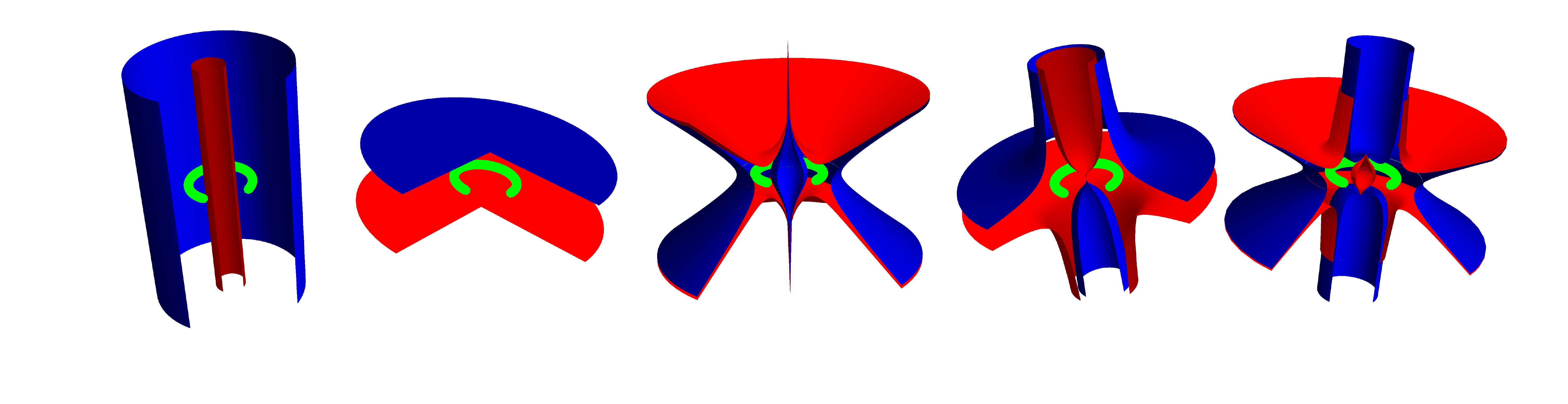}
\caption{\label{fig:SMult} Normal and skew 2n-pole magnets in cylindrical
	coordinates.
	Each figure shows magnetic (electric) field streamlines and poles' shape
	in transverse cross section.
	North (positive electrostatic potential) and
	south (negative electrostatic potential) poles are shown in red and blue
	and given by constant levels of
	$(\mathcal{B,A})_n^{\text{(e)}}=\mp\text{const}$ respectively,
	$\text{const} = 1$ for this example.
	Bottom row shows 3D models of sector magnets with
	$\theta = 3\pi/2$:
	skew S-dipole, normal S-dipole, skew S-quadrupole, normal S-quadrupole
	and skew S-sextupole from the left to the right respectively.
	Equilibrium orbit is shown in green color}
\end{figure*}

%------------------------------------------------------------------------------%
Finally, in order to define the set of functions for pure S-multipoles
(Table~\ref{tab:S-El}) we will define {\it sector harmonics}:
\begin{eqnarray*}
	\Cne(\rho,y)	&=&
	\sum_{k=0}^n \begin{pmatrix} n \\ k \end{pmatrix}
	\mathcal{F}_{n-k}(\rho)\,y^k\cos\frac{k\,\pi}{2},		\\
	\Cnm(\rho,y)	&=&
	\sum_{k=0}^n \begin{pmatrix} n \\ k \end{pmatrix}
	\frac{\mathcal{G}_{n-k}(\rho)}{\rho}\,y^k\cos\frac{k\,\pi}{2},	\\
	\Dne(\rho,y)	&=&
	\sum_{k=0}^n \begin{pmatrix} n \\ k \end{pmatrix}
	\mathcal{F}_{n-k}(\rho)\,y^k\sin\frac{k\,\pi}{2},		\\
	\Dnm(\rho,y)	&=&
	\sum_{k=0}^n \begin{pmatrix} n \\ k \end{pmatrix}
	\frac{\mathcal{G}_{n-k}(\rho)}{\rho}\,y^k\sin\frac{k\,\pi}{2}.
\end{eqnarray*}
obeying differential relations
\begin{eqnarray*}
n\,\{\mathcal{A,B}\}_{n-1}^{(\text{e}\,)} &=&
\pm \frac{\pd\,\{\mathcal{B,A}\}_n^{(\text{e})}}{\pd y} =
\frac{1}{\rho}
	\frac{\pd\left(\rho\,\{\mathcal{A,B}\}_n^{(\text{m})}\right)}
	{\pd\rho},							\\
n\,\{\mathcal{A,B}\}_{n-1}^{(\text{m})} &=&
\pm \frac{1}{\bcancel{\rho}}
\frac{\pd\,\left(\bcancel{\rho}\,\{\mathcal{B,A}\}_n^{(\text{m})}\right)}
	{\pd y} =
\frac{\pd\,\{\mathcal{A,B}\}_n^{(\text{e})}}{\pd \rho}.
\end{eqnarray*}
Figure~\ref{fig:SMult} shows the cross section of idealized multipole magnet's 
poles and corresponding fields.
First six members of spherical harmonics are listed in Table~\ref{tab:CD} and
exact expressions for potentials and fields in 
Appendix~\ref{secAP:NSMult}.

%------------------------------------------------------------------------------%
\begin{table}[hb!]
\caption{\label{tab:S-El}
	Formulas for the scalar potential, azimuthal component of the vector 
	potential and field components for ``pure normal and skew $2n$-poles in
	cylindrical coordinates.}
\begin{ruledtabular}
\begin{tabular}{p{0.45\columnwidth}|p{0.5\columnwidth}}
Normal	& Skew								\\\hline
									\\
[-0.2cm]
$\displaystyle\FIon = -  \Cno\frac{\Dne}{n!}$				&
$\displaystyle\FIun = -  \Cnu\frac{\Cne}{n!}$				\\
[0.2cm]
$\displaystyle\Aton = -  \Cno\frac{\Cnm}{n!}$				&
$\displaystyle\Atun =\quad\!\!\Cnu\frac{\Dnm}{n!}$			\\
[-0.3cm]
									\\\hline
									\\
[-0.2cm]
$\displaystyle\Fron =\quad\!\!\Cno
\frac{\mathcal{B}_{n-1}^{\text{(m)}}}{(n-1)!}$	&
$\displaystyle\Frun =\quad\!\!\Cnu
\frac{\mathcal{A}_{n-1}^{\text{(m)}}}{(n-1)!}$	\\
[0.2cm]
$\displaystyle\Fyon =\quad\!\!\Cno
\frac{\mathcal{A}_{n-1}^{\text{(e)}}}{(n-1)!}$	&
$\displaystyle\Fyun = -       \Cnu
\frac{\mathcal{B}_{n-1}^{\text{(e)}}}{(n-1)!}$
%------------------------------------------------------------------------------%
\end{tabular}
\end{ruledtabular}
\end{table}
%------------------------------------------------------------------------------%

% F- and G-functions tables ===================================================%
%==============================================================================%

%-------------------------------------------------------------------------------
\begin{table*}[p!]
\caption{\label{tab:Fn}First ten members of $\mathcal{F}-$functions.}
\begin{ruledtabular}
\begin{tabular}{lp{14cm}}
%-------------------------------------------------------------------------------
$n$	& $\Fn(\rho)$						\\[-0.3cm]
	&							\\\hline
	&							\\[-0.1cm]
%-------------------------------------------------------------------------------
0	&
$\displaystyle 1$						\\[0.4cm]
%-------------------------------------------------------------------------------
1	&
$\displaystyle \ln \rho$					\\[0.4cm]
%-------------------------------------------------------------------------------
2	&
$\displaystyle 
\frac{1}{2}\,(\rho^2-1) - \ln \rho$					
\\[0.3cm]
%-------------------------------------------------------------------------------
3	&
$\displaystyle \frac{3}{2} \left[
-(\rho^2-1)+(\rho^2+1)\ln \rho\phantom{\frac{1}{1}}\!\!\!\!
\right]$							\\[0.4cm]
%-------------------------------------------------------------------------------
4	&
$\displaystyle \,3\,\left[
\frac{1}{8}(\rho^4-1) + \frac{1}{2}(\rho^2-1)
	-\left( \rho^2 + \frac{1}{2} \right)\ln \rho
	\right]$						\\[0.4cm]
%-------------------------------------------------------------------------------
5	&
$\displaystyle\frac{15}{2}	\left[
	-\frac{3}{8}\left(\rho^4-1\right)
	+\left(\frac{1}{4}\,\rho^4+\rho^2+\frac{1}{4}\right)\ln \rho
		\right]$					\\[0.4cm]
%-------------------------------------------------------------------------------
6	&
$\displaystyle\frac{45}{4}	\left[
		  \frac{1}{36} \left(\rho^6-1\right)
		  +\frac{1}{2} \left(\rho^4-1\right)
		  -\frac{1}{4} \left(\rho^2-1\right)
	-\left(\frac{1}{2}\,\rho^4+\rho^2+\frac{1}{6}\right)\ln \rho
		\right]$					\\[0.4cm]
%-------------------------------------------------------------------------------
7	&
$\displaystyle\frac{315}{16}	\left[
			-\frac{11}{54}      \left(\rho^6-1\right)
			-\frac{1}{2}\,\rho^2\left(\rho^2-1\right)
			+\left\{
				\frac{1}{9}\left(\rho^6+1\right)+
				\rho^2\left(\rho^2+1\right)
			\right\}
			\ln \rho\right]$			\\[0.4cm]
%-------------------------------------------------------------------------------
8	&
$\displaystyle\frac{105}{4}	\left[
			\frac{1}{96} \left(\rho^8-1\right)
			+\frac{4}{9} \left(\rho^6-1\right)
			+\frac{3}{8} \left(\rho^4-1\right)
			-\frac{2}{3} \left(\rho^2-1\right)
	-\left(\frac{1}{3}\,\rho^6+
	       \frac{3}{2}\,\rho^4+
	       \rho^2+\frac{1}{12} \right)
			\ln \rho\right]$			\\[0.4cm]
%-------------------------------------------------------------------------------
9	&
$\displaystyle\frac{315}{8}	\left[
			-\frac{25}{192}     \left(\rho^8-1\right)
			-\frac{5}{6}\,\rho^2\left(\rho^4-1\right)
			+\left\{
				\frac{1}{16}+
				\rho^2\left(\frac{\rho^2}{2}+1\right)
				\left(\frac{1}{8}\,\rho^4+
				      \frac{7}{4}\,\rho^2+1\right)
		\right\}
		\ln \rho\right]$
%-------------------------------------------------------------------------------
\end{tabular}
\end{ruledtabular}
\end{table*}
%-------------------------------------------------------------------------------

%-------------------------------------------------------------------------------
\begin{table*}[p!]
\caption{\label{tab:Gn}First ten members of $\mathcal{G}-$functions.}
\begin{ruledtabular}
\begin{tabular}{lp{14cm}}
%-------------------------------------------------------------------------------
$n$	& $\Gn(\rho)$						\\[-0.3cm]
	&							\\\hline
	&							\\[-0.1cm]
%-------------------------------------------------------------------------------
0	&
$\displaystyle 1$						\\[0.4cm]
%-------------------------------------------------------------------------------
1	&
$\displaystyle 
\frac{1}{2}\,(\rho^2-1)$					\\[0.4cm]
%-------------------------------------------------------------------------------
2	&
$\displaystyle\,1\,\!\left[
	-\frac{1}{2}\,(\rho^2-1) + \rho^2\ln \rho
	\right]$						\\[0.4cm]
%-------------------------------------------------------------------------------
3	&
$\displaystyle \frac{3}{2} \left[
			\frac{1}{4}\,(\rho^4-1) -
			\rho^2 \ln \rho
	\right]$						\\[0.4cm]
%-------------------------------------------------------------------------------
4	&
$\displaystyle\,3\,\left[
	-\frac{5}{8}\,(\rho^4-1) + \frac{1}{2}(\rho^2-1)
	+\rho^2 \left(\frac{\rho^2}{2}+1\right) \ln \rho
	\right]$						\\[0.4cm]
%-------------------------------------------------------------------------------
5	&
$\displaystyle \frac{15}{4} \left[
		  \frac{1}{12}\,\left(\rho^6-1\right)
		  +\frac{3}{4}\,\rho^2\left(\rho^2-1\right)
	-\rho^2\left(\rho^2+1\right)\ln \rho
		\right]$					\\[0.4cm]
%-------------------------------------------------------------------------------
6	&
$\displaystyle \frac{45}{8} \left[
			-\frac{5}{9}\,\left(\rho^6-1\right)
			-\frac{1}{2}\,\left(\rho^4-1\right)
			+\left(\rho^2-1\right)			
			+\rho^2\left(
				\frac{1}{3}\,\rho^4+2\,\rho^2+1
			\right)\ln \rho
		\right]$					\\[0.4cm]
%-------------------------------------------------------------------------------
7	&
$\displaystyle \frac{105}{16} \left[
			\frac{1}{24}\,   \left(\rho^8-1\right)
			+\frac{7}{6}\,\rho^2\left(\rho^4-1\right)
	-\left(\rho^6+3\,\rho^4+\rho^2\right)
			\ln \rho
			\right]$				\\[0.4cm]
%-------------------------------------------------------------------------------
8	&
$\displaystyle\,\,\frac{35}{4}\,\left[
			-\frac{47}{96}\,\left(\rho^8-1\right)
			-2            \,\left(\rho^6-1\right)
			+\frac{9}{8}  \,\left(\rho^4-1\right)
			+\frac{4}{3}  \,\left(\rho^2-1\right)
			+\rho^2\left(
\frac{1}{4}\,\rho^6+3\,\rho^4+\frac{9}{2}\,\rho^2+1
			\right)
			\ln \rho
		\right]$					\\[0.4cm]
%-------------------------------------------------------------------------------
9	&
$\displaystyle \frac{315}{32} \left[
			 \frac{1} {40}\,   \left(\rho^{10}-1\right)
			+\frac{35}{24}\, \rho^2\left(\rho^6   -1\right)
			+\frac{5} {2}\,  \rho^4\left(\rho^2   -1\right)
			-\left(
			\rho^8+6\,\rho^6+6\,\rho^4+\rho^2
			\right) \ln \rho
		\right]$
%-------------------------------------------------------------------------------
\end{tabular}
\end{ruledtabular}
\end{table*}
%-------------------------------------------------------------------------------

% F- and G-functions plot =====================================================%
%==============================================================================%

%------------------------------------------------------------------------------%
\begin{figure*}[p!]
\includegraphics[width=0.8\linewidth]{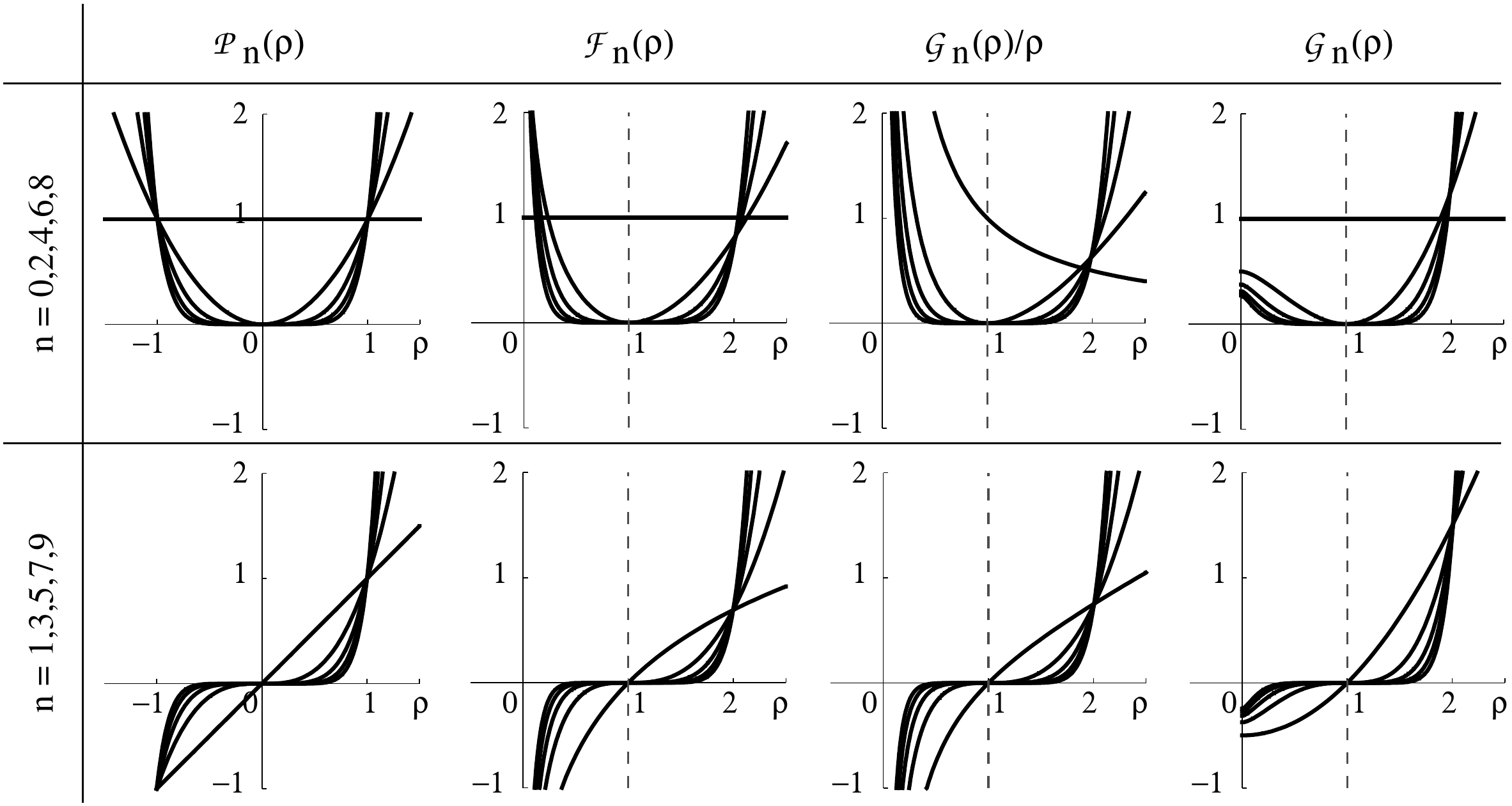}
\caption{\label{fig:FG} First five even (top row) and odd (bottom row) members
	of regular polynomials $\mathcal{P}_n = \rho^n$, $\Fn(\rho)$,
	$\frac{\Gn(\rho)}{\rho}$ and $\Gn(\rho)$ functions from the left to the 
	right respectively.}
\end{figure*}
%------------------------------------------------------------------------------%

% S-multipoles plot ===========================================================%
%==============================================================================%

%------------------------------------------------------------------------------%
%------------------------------------------------------------------------------%

% Sector harmonics ============================================================%
%==============================================================================%

%------------------------------------------------------------------------------%
\begin{table*}[p!]
\caption{\label{tab:CD}Sector harmonics.}
\begin{ruledtabular}
\begin{tabular}{lll}
%------------------------------------------------------------------------------%
	&
n	&							\\\hline
								\\[-0.3cm]
%------------------------------------------------------------------------------%
$\Cne$	&
0	& $\displaystyle
		1       \phantom{\left[\frac{1}{1}\right]}$	\\
&1	& $\displaystyle
		\ln \rho\phantom{\left[\frac{1}{1}\right]}$	\\[0.3cm]
&2	& $\displaystyle
	\left[
		\frac{\rho^2-1}{2}-y^2
	\right] - \ln \rho$					\\[0.3cm]
&3	& $\displaystyle
	\left[
		-3\,\frac{\rho^2-1}{2}
	\right]+
	3\,\left(
		\frac{\rho^2+1}{2}-y^2
	\right)\ln \rho$					\\[0.3cm]
&4	& $\displaystyle
	\left[
		 \frac{3(\rho^4+4\,\rho^2-5)}{8}
		-6\,\frac{\rho^2-1}{2}\,y^2
		+y^4
	\right]
	-3\,\left(
		\frac{1}{2}+\rho^2-2\,y^2
	\right)\,\ln \rho$					\\[0.3cm]
&5	& $\displaystyle
	\left[
		-5 \,\frac{9\,(\rho^4-1)}{16}
		+5\times 6\frac{\rho^2-1}{2}\,y^2
	\right]
	+5\,\left(
		\frac{3(\rho^4+4\,\rho^2+1)}{8}
		-6\,\frac{\rho^2+1}{2}\,y^2
		+y^4
	\right)\ln \rho$					\\[-0.3cm]
								\\\hline
								\\[-0.3cm]
%------------------------------------------------------------------------------%
$\Cnm$	&
0	& $\displaystyle \frac{1}{\rho}
	\left\{
		1 \phantom{\frac{1}{1}}\!\!\!\!
	\right\}$						\\[0.3cm]
&1	& $\displaystyle \frac{1}{\rho}\left\{
	\left[ \frac{\rho^2-1}{2} \right]
	\right\}$						\\[0.3cm]
&2	& $\displaystyle \frac{1}{\rho}\left\{
	\left[-\frac{\rho^2-1}{2} - y^2\right]
	+\rho^2\ln \rho
	\right\}$						\\[0.3cm]
&3	& $\displaystyle \frac{1}{\rho}\left\{
	\left[
		 \frac{3(\rho^2+1)}{4}\frac{\rho^2-1}{2}
		-3\,\frac{\rho^2-1}{2}\,y^2
	\right]
	- \frac{3}{2}\,\rho^2\ln \rho
	\right\}$						\\[0.3cm]
&4	& $\displaystyle \frac{1}{\rho}\left\{
	\left[
		-\frac{3(5\,\rho^4-4\,\rho^2-1)}{8}
		+6\,\frac{\rho^2-1}{2}\,y^2
		+y^4
	\right]
	+ \frac{3(2+\rho^2-4\,y^2)}{2}\,\rho^2\ln \rho
	\right\}$						\\[0.3cm]
&5	& $\displaystyle \frac{1}{\rho}\left\{
	\left[
		 \frac{5(\rho^4+10\,\rho^2+1)}{8}\frac{\rho^2-1}{2}
		-10\,\frac{3(\rho^2+1)}{4}\frac{\rho^2-1}{2}\,y^2
		+5\,\frac{\rho^2-1}{2}\,y^4
	\right]
	- \frac{15(1+\rho^2-4\,y^2)}{4}\,\rho^2\ln \rho
	\right\}$						\\[-0.3cm]
								\\\hline
								\\[-0.3cm]
%------------------------------------------------------------------------------%
$\Dne$	&
0	& $\displaystyle 0 \phantom{\frac{1}{1}}$		\\[0.3cm]
&1	& $\displaystyle y\left\{
	\!\!\!\!\!\!\!\!\!\!\!\phantom{\left[ \frac{1}{1} \right]}
	1
	\right\}$						\\[0.3cm]
&2	& $\displaystyle y\left\{
	\!\!\!\!\!\!\!\!\!\!\!\phantom{\left[ \frac{1}{1} \right]}
		2\,\ln \rho
	\right\}$						\\[0.3cm]
&3	& $\displaystyle y\left\{
	\left[
		3\,\frac{\rho^2-1}{2} - y^2
	\right]
	-3\,\ln \rho
	\right\}$						\\[0.3cm]
&4	& $\displaystyle y\left\{
	\left[
		-12\,\frac{\rho^2-1}{2}
	\right]
	+4\,\left(
		3\,\frac{\rho^2+1}{2}-y^2
	\right)\ln \rho
	\right\}$						\\[0.3cm]
&5	& $\displaystyle y\left\{
	\left[
		 5 \,\frac{3\,(\rho^4+4\,\rho^2-5)}{8}
		-10\,\frac{\rho^2-1}{2}\,y^2
		+y^4
	\right]
	-5\left(
		\frac{3}{2}+3\,\rho^2-2\,y^2
	\right)\ln \rho
	\right\}$						\\[-0.3cm]
								\\\hline
								\\[-0.3cm]
%------------------------------------------------------------------------------%
$\Dnm$	&
0	& $\,\displaystyle 0 \phantom{\frac{1}{1}}$		\\[0.3cm]
&1	& $\displaystyle \frac{y}{\rho}\left\{
	\!\!\!\!\!\!\!\!\!\!\!\phantom{\left[ \frac{1}{1} \right]}
	1
	\right\} $						\\[0.3cm]
&2	& $\displaystyle \frac{y}{\rho}\left\{
	\left[2\,\frac{\rho^2-1}{2}\right]
	\right\}$		\\[0.3cm]
&3	& $\displaystyle \frac{y}{\rho}\left\{
	\left[
		 -3\,\frac{\rho^2-1}{2}- y^2
	\right]
	+ 3\,\rho^2\ln \rho
	\right\}$	\\[0.3cm]
&4	& $\displaystyle \frac{y}{\rho}\left\{
	\left[
		 4\,\frac{3(\rho^2+1)}{4}\frac{\rho^2-1}{2}
		-4\,\frac{\rho^2-1}{2}\,y^2
	\right]
	- 6\,\rho^2\ln \rho
	\right\}$	\\[0.3cm]
&5	& $\displaystyle \frac{y}{\rho}\left\{
	\left[
		-5\,\frac{3(5\,\rho^4-4\,\rho^2-1)}{8}
		+10\,\frac{\rho^2-1}{2}\,y^2
		+y^4
	\right]
	+5\,\left(3+\frac{3}{2}\,\rho^2-2\,y^2\right)\rho^2\ln \rho
	\right\}$
%------------------------------------------------------------------------------%
\end{tabular}
\end{ruledtabular}
\end{table*}
%------------------------------------------------------------------------------%

% DIFFERENCE EQUATIONS ========================================================%
%==============================================================================%
%==============================================================================%
%==============================================================================%
%==============================================================================%
\subsection{\label{sec:Rec}Recurrence equations in sector coordinates}

%------------------------------------------------------------------------------%
An alternative approach to find expansions for potentials is to use general
power series ansatz.
In Cartesian coordinates the use of
\[
	\Phi = -\sum_{m,n \geq 0}^\infty V_{m,n}\frac{x^m}{m!}\frac{y^n}{n!}
\]
gives the recurrence relation
\[
	V_{m+2,n} + V_{m,n+2} = 0.
\]
This equation immediately defines all coefficients, and up to a common factor,
as easy to see, coincides with harmonic homogeneous polynomials $\An$ and $\Bn$.

%------------------------------------------------------------------------------%
In sector coordinates, the same substitution for $\Phi$, and
\[
	\At = -\sum_{m,n \geq 0}^\infty
		\frac{1}{1+x}V_{m,n}\frac{x^m}{m!}\frac{y^n}{n!}
\]
substitution for longitudinal component of the vector potential gives two new
recurrences, respectively
\[
	V_{m+2,n} + V_{m,n+2} =
		-(m \pm 1)\,V_{m+1.n} - m\,V_{m-1,n+2}.
\]
The detailed approach on how to treat these equations can be found for example
in [Wiedemann].
In order to solve these recurrences, one can look for a solution where each 
term can be expressed in a form
\[
	V_{i,j} = V_{i,j}^*	+ V_{i,j}^{(i+j-1)}
				+ V_{i,j}^{(i+j-2)}
				+ V_{i,j}^{(i+j-3)}
				+ \ldots
\]
where starred variables are the ``design'' terms given by pure multipole fields
and thus satisfying
\[
	V_{m+2,n}^* + V_{m,n+2}^* \equiv 0.
\]
Other coefficients $V_{i,j}^{(k)}$ are terms induced by lower $k$-th order pure
multipoles due to recurrence.
Thus in order to find an expression for a particular $2n$-pole we will start 
the recurrence form the $n$-th order assuming that
\[
	V_{n,0}   = -V_{n-2,2} = \ldots
		\quad\text{or}\quad
	V_{n-1,1} = -V_{n-3,3} = \ldots
\]
for normal and skew elements.
Then we will start exploiting the recurrence where all terms in the form
$V_{i,j}^{(n)}$ for $i+j>n$ are subject to be determined.

%------------------------------------------------------------------------------%
This approach has two major disadvantages.
At first, in order to use the result on will have to truncate a recurrence.
As a result the potentials representing magnets do not satisfies the Laplace
equation anymore.
This is a strong assumption which violate the ``physics'' and should be avoided.
While potentials can be approximated with any precision by keeping an 
appropriate
number of terms, there is another issue.
At second, at each new order when solving the recurrence one will find that an
arbitrary constant $\alpha_i\in(0;1)$ should be introduced since the system is
undetermined.
An additional assumption $(A_s,\Phi)\left.\right|_{x=0}\propto y^n$ allows to
truncate or summate the series.
The resulting solutions coincide with the one obtained above.

%==============================================================================%
%==============================================================================%
%==============================================================================%
\section{Summary}

%------------------------------------------------------------------------------%
The scalar and vector Laplace's equations for static transverse electromagnetic
fields in curvilinear orthogonal coordinates with zero and constant curvature 
are solved.
In Cartesian coordinates these solutions are well known harmonic homogeneous 
polynomials of two variables.
The set of solutions in cylindrical coordinates named sector harmonics, and
should not be confused with cylindrical harmonics where $\rho$-dependent term 
is given by Bessel functions which occasionally are also called cylindrical 
harmonics.
In contrast, the radial part is given by the set of introduced McMillan radial
harmonics, independently introduced by E.M.~McMillan in his ``forgotten'' 
article, and adjoint radial harmonics also described in this work.
The feature of sector harmonics that when expanded around equilibrium orbit they
resemble solution in Cartesian geometry.
Compared to the traditional approach, widely used in accelerator community,
of the use of recurrences based on general power series ansatz, this 
set of functions has two major advantages.
It do not require any truncation and is exactly satisfying Laplace equation, 
and,
provides a well defined full basis of functions which can be related to any 
field by its expansion in radial or vertical planes, see Table~\ref{tab:CnS}.
Including the model Hamiltonians for $t$- and $s$-representations, where no 
assumptions but the field symmetry has been used, one can construct numerical
scheme integrating equations of motion.
Thus I would like to suggest the set of sector harmonics as a new basis for 
description and design of any sector magnets with translational symmetry along
azimuthal coordinate.

%------------------------------------------------------------------------------%
\begin{table*}[t!]
\caption{\label{tab:CnS}
	Relationship between coefficients determining the strength of ``pure'' 
	normal and skew S-multipoles and power series expansion of field in 
	radial and vertical planes on equilibrium orbit.}
\begin{ruledtabular}
\begin{tabular}{p{0.2cm}p{0.2cm}p{1.5cm}l}
%------------------------------------------------------------------------------%
&$n$ & $x=0$	& $y=0$			\\\hline
%------------------------------------------------------------------------------%
$\Cno$
%------------------------------------------------------------------------------%
&1 &
$	F_y$				&
$	F_y$				\\[0.2cm]
%------------------------------------------------------------------------------%
&2 &
$	\pd_y\,F_x$			&
$	\pd_x\,F_y$			\\[0.2cm]
%------------------------------------------------------------------------------%
&3 &
$-	\pd_y^2\,F_y$			&
$	 \pd_x^2\,F_y
	+\pd_x  \,F_y$			\\[0.2cm]
%------------------------------------------------------------------------------%
&4 &
$-	\pd_y^3\,F_x$			&
$	 \pd_x^3\,F_y
	+\pd_x^2\,F_y
	-\pd_x  \,F_y$			\\[0.2cm]
%------------------------------------------------------------------------------%
&5 &
$	  \pd_y^4\,F_y$			&
$	    \pd_x^4\,F_y
	+2\,\pd_x^3\,F_y
	-   \pd_x^2\,F_y
	+   \pd_x  \,F_y$		\\[0.2cm]
%------------------------------------------------------------------------------%
&6 &
$	\pd_y^5\,F_x$			&
$	    \pd_x^5\,F_y
	+2\,\pd_x^4\,F_y
	-3\,\pd_x^3\,F_y
	+3\,\pd_x^2\,F_y
	-3\,\pd_x  \,F_y$		\\[0.2cm]
%------------------------------------------------------------------------------%
&7 &
$-	\pd_y^6\,F_y$			&
$	    \pd_x^6\,F_y
	+3\,\pd_x^5\,F_y
	-3\,\pd_x^4\,F_y
	+6\,\pd_x^3\,F_y
	-9\,\pd_x^2\,F_y
	+9\,\pd_x  \,F_y$		\\[0.2cm]
%------------------------------------------------------------------------------%
&8 &
$-	\pd_y^7\,F_x$			&
$	     \pd_x^7\,F_y
	+ 3\,\pd_x^6\,F_y
	- 6\,\pd_x^5\,F_y
	+12\,\pd_x^4\,F_y
	-27\,\pd_x^3\,F_y
	+45\,\pd_x^2\,F_y
	-45\,\pd_x  \,F_y$		\\[0.2cm]
%------------------------------------------------------------------------------%
&9 &
$	\pd_y^8\,F_y$			&
$	      \pd_x^8\,F_y
	+  4\,\pd_x^7\,F_y
	-  6\,\pd_x^6\,F_y
	+ 18\,\pd_x^5\,F_y
	- 51\,\pd_x^4\,F_y
	+126\,\pd_x^3\,F_y
	-225\,\pd_x^2\,F_y
	+225\,\pd_x  \,F_y$		\\[-0.2cm]
					\\\hline
%------------------------------------------------------------------------------%
$\Cnu$
%------------------------------------------------------------------------------%
&1 &
$	F_x$				&
$	F_x$				\\[0.1cm]
%------------------------------------------------------------------------------%
&2 &
$-	\pd_y\,F_y$			&
$	 \pd_x\,F_x
	+       F_x$			\\[0.1cm]
%------------------------------------------------------------------------------%
&3 &
$-	\pd_y^2\,F_x$			&
$	 \pd_x^2\,F_x
	+\pd_x  \,F_x
	-         F_x$			\\[0.1cm]
%------------------------------------------------------------------------------%
&4 &
$	\pd_y^3\,F_y$			&
$	    \pd_x^3\,F_x
	+2\,\pd_x^2\,F_x
	-   \pd_x  \,F_x
	+            F_x$		\\[0.1cm]
%------------------------------------------------------------------------------%
&5 &
$	\pd_y^4\,F_x$			&
$	    \pd_x^4\,F_x
	+2\,\pd_x^3\,F_x
	-3\,\pd_x^2\,F_x
	+3\,\pd_x  \,F_x
	-3\,         F_x$		\\[0.1cm]
%------------------------------------------------------------------------------%
&6 &
$-	\pd_y^5\,F_y$			&
$	    \pd_x^5\,F_x
	+3\,\pd_x^4\,F_x
	-3\,\pd_x^3\,F_x
	+6\,\pd_x^2\,F_x
	-9\,\pd_x  \,F_x
	+9\,         F_x$		\\[0.1cm]
%------------------------------------------------------------------------------%
&7 &
$-	\pd_y^6\,F_x$			&
$	     \pd_x^6\,F_x
	+ 3\,\pd_x^5\,F_x
	- 6\,\pd_x^4\,F_x
	+12\,\pd_x^3\,F_x
	-27\,\pd_x^2\,F_x
	+45\,\pd_x  \,F_x
	-45\,         F_x$		\\[0.1cm]
%------------------------------------------------------------------------------%
&8 &
$	\pd_y^7\,F_y$			&
$	      \pd_x^7\,F_x
	+  4\,\pd_x^6\,F_x
	-  6\,\pd_x^5\,F_x
	+ 18\,\pd_x^4\,F_x
	- 51\,\pd_x^3\,F_x
	+126\,\pd_x^2\,F_x
	-225\,\pd_x  \,F_x
	+225\,         F_x$		\\[0.1cm]
%------------------------------------------------------------------------------%
&9 &
$	\pd_y^8\,F_x$			&
$	       \pd_x^8\,F_x
	+  4 \,\pd_x^7\,F_x
	- 10 \,\pd_x^6\,F_x
	+ 30 \,\pd_x^5\,F_x
	-105 \,\pd_x^4\,F_x
	+330 \,\pd_x^3\,F_x
	-855 \,\pd_x^2\,F_x
	+1575\,\pd_x  \,F_x
	-1575\,         F_x$		\\[0.1cm]
%------------------------------------------------------------------------------%
\end{tabular}
\end{ruledtabular}
\end{table*}
%------------------------------------------------------------------------------%

%==============================================================================%
%==============================================================================%
%==============================================================================%
\begin{acknowledgments}
The author would like to thank
{\bf Leo~Michelotti},
{\bf Eric~Stern}
and
{\bf James~F.~Amundson}
for their discussions and valuable input.
{\bf Alexey~Burov} for encouraging to find full family of solutions.
{\bf Valeri~Lebedev} whose solution for electrostatic quadrupole led me to
generalization, just as in the case with {\bf E.~M.~McMillan} and
{\bf F.~Krienen}.
And, of course, {\bf Sergei~Nagaitsev} who brought back to life original
unknown McMillan's article which helped me with symmetric description of
electromagnetic fields.
\end{acknowledgments}

%==============================================================================%
%==============================================================================%
%==============================================================================%
%==============================================================================%
%==============================================================================%
\appendix

%==============================================================================%
%==============================================================================%
%==============================================================================%
\section{\label{secAP:NSMult}R- ans S- multipoles. Exact expressions.}

%-------------------------------------------------------------------------------
The scalar potentials, longitudinal component of vector potential, and field 
components for pure R- and S-multipoles up to fifth order are listed in 
Table~\ref{tab:AzPhi}--\ref{tab:FxFy} and Tables~\ref{tab:AtPhi}--\ref{tab:FrFy}
respectively.
%==============================================================================%
%==============================================================================%
%==============================================================================%
\section{\label{secAP:Taylor}Taylor polynomials of $\Fn$ and $\Gn$.}

%-------------------------------------------------------------------------------
The first ten terms of Maclaurin series of $\Fn(x)$, $\Gn(x)$ and
$\displaystyle \frac{\Gn(x)}{1+x}$ are listed in Table~\ref{tab:FGTaylor}.

% R-multipole POTENTIALS ======================================================%
%==============================================================================%

%------------------------------------------------------------------------------%
\begin{table*}[p!]
\caption{\label{tab:AzPhi}Longitudinal component of the vector potential
	and scalar potential for pure normal and skew R-multipoles.}
\begin{ruledtabular}
\begin{tabular}{p{1cm}p{3cm}p{6cm}p{6cm}}
%------------------------------------------------------------------------------%
$n$	&		& $A_z$		& $\Phi$		\\\hline
								\\[-0.3cm]
%------------------------------------------------------------------------------%
0	& calibration	&
$\displaystyle-\,\overline{C}_{0}$				&
$\,\,\,\,\,\,0$							\\[0.3cm]
1	& normal dipole	&
$\displaystyle-\frac{1}{1!}(x)\,\overline{C}_{1}$		&
$\displaystyle-\frac{1}{1!}(y)\,\overline{C}_{1}$		\\[0.3cm]
2	& normal quadrupole	&
$\displaystyle-\frac{1}{2!}
	(x^2-y^2)
	\,\overline{C}_{2}$					&
$\displaystyle-\frac{1}{2!}
	(2\,x\,y)
	\,\overline{C}_{2}$					\\[0.3cm]
3	& normal sextupole	&
$\displaystyle-\frac{1}{3!}
	(x^3-3\,x\,y^2)
	\,\overline{C}_{3}$					&
$\displaystyle-\frac{1}{3!}
	(3\,x^2y-y^3)
	\,\overline{C}_{3}$					\\[0.3cm]
4	& normal octupole	&
$\displaystyle-\frac{1}{4!}
	(x^4-6\,x^2y^2+y^4)
	\,\overline{C}_{4}$					&
$\displaystyle-\frac{1}{4!}
	(4\,x^3y-4\,x\,y^3)
	\,\overline{C}_{4}$					\\[0.3cm]
5	& normal decapole	&
$\displaystyle-\frac{1}{5!}
	(x^5-10\,x^3y^2+5\,x\,y^4)
	\,\overline{C}_{5}$					&
$\displaystyle-\frac{1}{5!}
	(5\,x^4y-10\,x^2y^3+y^5)
	\,\overline{C}_{5}$					\\[-0.3cm]
								\\\hline
								\\[-0.3cm]
%------------------------------------------------------------------------------%
0	& calibration	&
$\,\,\,\,\,0$							&
$\,\,\,\,\,\underline{C}_{\,0}$					\\[0.3cm]
1	& skew dipole	&
$\,\,\,\,\displaystyle \frac{1}{1!}(y)\,\underline{C}_{\,1}$	&
$\displaystyle-\frac{1}{1!}(x)\,\underline{C}_{\,1}$		\\[0.3cm]
2	& skew quadrupole	&
$\,\,\,\,\displaystyle \frac{1}{2!}
	(2\,x\,y)
	\,\underline{C}_{\,2}$					&
$\displaystyle-\frac{1}{2!}
	(x^2-y^2)
	\,\underline{C}_{\,2}$					\\[0.3cm]
3	& skew sextupole	&
$\,\,\,\,\displaystyle \frac{1}{3!}
	(3\,x^2y-y^3)
	\,\underline{C}_{\,3}$					&
$\displaystyle-\frac{1}{3!}
	(x^3-3\,x\,y^2)
	\,\underline{C}_{\,3}$					\\[0.3cm]
4	& skew octupole	&
$\,\,\,\,\displaystyle \frac{1}{4!}
	(4\,x^3y-4\,x\,y^3)
	\,\underline{C}_{\,4}$					&
$\displaystyle-\frac{1}{4!}
	(x^4-6\,x^2y^2+y^4)
	\,\underline{C}_{\,4}$					\\[0.3cm]
5	& skew decapole	&
$\,\,\,\,\displaystyle \frac{1}{5!}
	(5\,x^4y-10\,x^2y^3+y^5)
	\,\underline{C}_{\,5}$					&
$\displaystyle-\frac{1}{5!}
	(x^5-10\,x^3y^2+5\,x\,y^4)
	\,\underline{C}_{\,5}$
%------------------------------------------------------------------------------%
\end{tabular}
\end{ruledtabular}
\end{table*}
%------------------------------------------------------------------------------%

% R-multipole FIELDS ==========================================================%
%==============================================================================%

%------------------------------------------------------------------------------%
\begin{table*}[p!]
\caption{\label{tab:FxFy}Horizontal and vertical components of pure 
	normal and skew R-multipole magnets' field.}
\begin{ruledtabular}
\begin{tabular}{p{1cm}p{3cm}p{6cm}p{6cm}}
%------------------------------------------------------------------------------%
$n$	&		& $F_x$	& $F_y$				\\\hline
								\\[-0.3cm]
%------------------------------------------------------------------------------%
0	& calibration	&
---								&
$\,\,\,\,$---							\\[0.3cm]
1	& normal dipole	&
$\,0$								&
$\,\,\,\,\,\overline{C}_{1}$					\\[0.3cm]
2	& normal quadrupole	&
$\displaystyle \frac{1}{1!}
	(y)
	\,\overline{C}_{2}$					&
$\,\,\,\,\displaystyle \frac{1}{1!}
	(x)
	\,\overline{C}_{2}$					\\[0.3cm]
3	& normal sextupole	&
$\displaystyle \frac{1}{2!}
	(2\,x\,y)
	\,\overline{C}_{3}$					&
$\,\,\,\,\displaystyle \frac{1}{2!}
	(x^2-y^2)
	\,\overline{C}_{3}$					\\[0.3cm]
4	& normal octupole	&
$\displaystyle \frac{1}{3!}
	(3\,x^2y-y^3)
	\,\overline{C}_{4}$					&
$\,\,\,\,\displaystyle \frac{1}{3!}
	(x^3-3\,x\,y^2)
	\,\overline{C}_{4}$					\\[0.3cm]
5	& normal decapole	&
$\displaystyle \frac{1}{4!}
	(4\,x^3y-4\,x\,y^3)
	\,\overline{C}_{5}$		&
$\,\,\,\,\displaystyle \frac{1}{4!}
	(x^4-6\,x^2y^2+y^4)
	\,\overline{C}_{5}$					\\[-0.3cm]
								\\\hline
								\\[-0.3cm]
%------------------------------------------------------------------------------%
0	& calibration	&
---								&
$\,\,\,\,\,$---							\\[0.3cm]
1	& skew dipole	&
$\,\underline{C}_{\,1}$						&
$\,\,\,\,\,\,0$							\\[0.3cm]
2	& skew quadrupole	&
$\displaystyle \frac{1}{1!}
	(x)
	\,\underline{C}_{\,2}$		&
$\displaystyle-\frac{1}{1!}
	(y)
	\,\underline{C}_{\,2}$		\\[0.3cm]
3	& skew sextupole	&
$\displaystyle \frac{1}{2!}
	(x^2-y^2)
	\,\underline{C}_{\,3}$		&
$\displaystyle-\frac{1}{2!}
	(2\,x\,y)
	\,\underline{C}_{\,3}$		\\[0.3cm]
4	& skew octupole	&
$\displaystyle \frac{1}{3!}
	(x^3-3\,x\,y^2)
	\,\underline{C}_{\,4}$		&
$\displaystyle-\frac{1}{3!}
	(3\,x^2y-y^3)
	\,\underline{C}_{\,4}$		\\[0.3cm]
5	& skew decapole	&
$\displaystyle \frac{1}{4!}
	(x^4-6\,x^2y^2+y^4)
	\,\underline{C}_{\,5}$		&
$\displaystyle-\frac{1}{4!}
	(4\,x^3y-4\,x\,y^3)
	\,\underline{C}_{\,5}$	
%------------------------------------------------------------------------------%
\end{tabular}
\end{ruledtabular}
\end{table*}
%------------------------------------------------------------------------------%

% S-multipole POTENTIALS ======================================================%
%==============================================================================%

%------------------------------------------------------------------------------%
\begin{table*}[t!]
\caption{\label{tab:AtPhi} Azimuthal component of the vector potential
	and scalar potential for ``pure'' normal and skew S-multipoles.}
\begin{ruledtabular}
\begin{tabular}{lll}
%------------------------------------------------------------------------------%
	&
n	&							\\\hline
								\\[-0.3cm]
%------------------------------------------------------------------------------%
$\Aton$	&
0	& $\displaystyle -\frac{1}{0!}\frac{1}{\rho}
	\left\{
		1 \phantom{\frac{1}{1}}\!\!\!\!
	\right\}\,\overline{C}_{0}$				\\[0.3cm]
&1	& $\displaystyle -\frac{1}{1!}\frac{1}{\rho}\left\{
	\left[ \frac{\rho^2-1}{2} \right]
	\right\}\,\overline{C}_{1}$				\\[0.3cm]
&2	& $\displaystyle -\frac{1}{2!}\frac{1}{\rho}\left\{
	\left[-\frac{\rho^2-1}{2} - y^2\right]
	+\rho^2\ln \rho
	\right\}
	\,\overline{C}_{2}$					\\[0.3cm]
&3	& $\displaystyle -\frac{1}{3!}\frac{1}{\rho}\left\{
	\left[
		 \frac{3(\rho^2+1)}{4}\frac{\rho^2-1}{2}
		-3\,\frac{\rho^2-1}{2}\,y^2
	\right]
	- \frac{3}{2}\,\rho^2\ln \rho
	\right\}
	\,\overline{C}_{3}$					\\[0.3cm]
&4	& $\displaystyle -\frac{1}{4!}\frac{1}{\rho}\left\{
	\left[
		-\frac{3(5\,\rho^4-4\,\rho^2-1)}{8}
		+6\,\frac{\rho^2-1}{2}\,y^2
		+y^4
	\right]
	+ \frac{3(2+\rho^2-4\,y^2)}{2}\,\rho^2\ln \rho
	\right\}
	\,\overline{C}_{4}$					\\[0.3cm]
&5	& $\displaystyle -\frac{1}{5!}\frac{1}{\rho}\left\{
	\left[
		 \frac{5(\rho^4+10\,\rho^2+1)}{8}\frac{\rho^2-1}{2}
		-10\,\frac{3(\rho^2+1)}{4}\frac{\rho^2-1}{2}\,y^2
		+5\,\frac{\rho^2-1}{2}\,y^4
	\right]
	- \frac{15(1+\rho^2-4\,y^2)}{4}\,\rho^2\ln \rho
	\right\}
	\,\overline{C}_{5}$					\\[-0.3cm]
								\\\hline
								\\[-0.3cm]
%------------------------------------------------------------------------------%
$\FIon$	&
0	& $\quad\displaystyle 0 \phantom{\frac{1}{1}}$		\\[0.3cm]
&1	& $\displaystyle -\frac{1}{1!}\,y\left\{
	\!\!\!\!\!\!\!\!\!\!\!\phantom{\left[ \frac{1}{1} \right]}
	1
	\right\}\,\overline{C}_{1}$				\\[0.3cm]
&2	& $\displaystyle -\frac{1}{2!}\,y\left\{
	\!\!\!\!\!\!\!\!\!\!\!\phantom{\left[ \frac{1}{1} \right]}
		2\,\ln \rho
	\right\}
	\,\overline{C}_{2}$					\\[0.3cm]
&3	& $\displaystyle -\frac{1}{3!}\,y\left\{
	\left[
		3\,\frac{\rho^2-1}{2} - y^2
	\right]
	-3\,\ln \rho
	\right\}
	\,\overline{C}_{3}$					\\[0.3cm]
&4	& $\displaystyle -\frac{1}{4!}\,y\left\{
	\left[
		-12\,\frac{\rho^2-1}{2}
	\right]
	+4\,\left(
		3\,\frac{\rho^2+1}{2}-y^2
	\right)\ln \rho
	\right\}
	\,\overline{C}_{4}$					\\[0.3cm]
&5	& $\displaystyle -\frac{1}{5!}\,y\left\{
	\left[
		 5 \,\frac{3\,(\rho^4+4\,\rho^2-5)}{8}
		-10\,\frac{\rho^2-1}{2}\,y^2
		+y^4
	\right]
	-5\left(
		\frac{3}{2}+3\,\rho^2-2\,y^2
	\right)\ln \rho
	\right\}
	\,\overline{C}_{5}$					\\[-0.3cm]
								\\\hline
								\\[-0.3cm]
%------------------------------------------------------------------------------%
$\Atun$	&
0	& $\quad\!\,\displaystyle 0 \phantom{\frac{1}{1}}$	\\[0.3cm]
&1	& $\quad\!\displaystyle \frac{1}{1!}\frac{y}{\rho}\left\{
	\!\!\!\!\!\!\!\!\!\!\!\phantom{\left[ \frac{1}{1} \right]}
	1
	\right\} \,\underline{C}_{\,1}$				\\[0.3cm]
&2	& $\quad\!\displaystyle \frac{1}{2!}\frac{y}{\rho}\left\{
	\left[2\,\frac{\rho^2-1}{2}\right]
	\right\}
	\,\underline{C}_{\,2}$					\\[0.3cm]
&3	& $\quad\!\displaystyle \frac{1}{3!}\frac{y}{\rho}\left\{
	\left[
		 -3\,\frac{\rho^2-1}{2}- y^2
	\right]
	+ 3\,\rho^2\ln \rho
	\right\}
	\,\underline{C}_{\,3}$					\\[0.3cm]
&4	& $\quad\!\displaystyle \frac{1}{4!}\frac{y}{\rho}\left\{
	\left[
		 4\,\frac{3(\rho^2+1)}{4}\frac{\rho^2-1}{2}
		-4\,\frac{\rho^2-1}{2}\,y^2
	\right]
	- 6\,\rho^2\ln \rho
	\right\}
	\,\underline{C}_{\,4}$					\\[0.3cm]
&5	& $\quad\!\displaystyle \frac{1}{5!}\frac{y}{\rho}\left\{
	\left[
		-5\,\frac{3(5\,\rho^4-4\,\rho^2-1)}{8}
		+10\,\frac{\rho^2-1}{2}\,y^2
		+y^4
	\right]
	+5\,\left(3+\frac{3}{2}\,\rho^2-2\,y^2\right)\rho^2\ln \rho
	\right\}
	\,\underline{C}_{\,5}$					\\[-0.3cm]
								\\\hline
								\\[-0.3cm]
%------------------------------------------------------------------------------%
$\FIun$	&
0	& $\displaystyle-\frac{1}{0!}
	\left\{
		1\phantom{\frac{1}{1}}\!\!\!\!\!
	\right\}\,\underline{C}_{\,0}$				\\[0.3cm]
&1	& $\displaystyle -\frac{1}{1!}\left\{
	\!\!\!\!\!\!\!\!\!\!\!\phantom{\left[ \frac{1}{1} \right]}
		\ln \rho\phantom{\frac{1}{1}}\!\!\!\!
	\right\} \,\underline{C}_{\,1}$				\\[0.3cm]
&2	& $\displaystyle -\frac{1}{2!}\left\{
	\left[
		\frac{\rho^2-1}{2}-y^2
	\right] - \ln \rho
	\right\}
	\,\underline{C}_{\,2}$					\\[0.3cm]
&3	& $\displaystyle -\frac{1}{3!}\left\{
	\left[
		-3\,\frac{\rho^2-1}{2}
	\right]+
	3\,\left(
		\frac{\rho^2+1}{2}-y^2
	\right)\ln \rho
	\right\}
	\,\underline{C}_{\,3}$					\\[0.3cm]
&4	& $\displaystyle -\frac{1}{4!}\left\{
	\left[
		 \frac{3(\rho^4+4\,\rho^2-5)}{8}
		-6\,\frac{\rho^2-1}{2}\,y^2
		+y^4
	\right]
	-3\,\left(
		\frac{1}{2}+\rho^2-2\,y^2
	\right)\,\ln \rho
	\right\}
	\,\underline{C}_{\,4}$					\\[0.3cm]
&5	& $\displaystyle -\frac{1}{5!}\left\{
	\left[
		-5 \,\frac{9\,(\rho^4-1)}{16}
		+5\times 6\frac{\rho^2-1}{2}\,y^2
	\right]
	+5\,\left(
		\frac{3(\rho^4+4\,\rho^2+1)}{8}
		-6\,\frac{\rho^2+1}{2}\,y^2
		+y^4
	\right)\ln \rho
	\right\}
	\,\underline{C}_{\,5}$
%------------------------------------------------------------------------------%
\end{tabular}
\end{ruledtabular}
\end{table*}
%------------------------------------------------------------------------------%

% S-multipole FIELDS ==========================================================%
%==============================================================================%

%------------------------------------------------------------------------------%
\begin{table*}[t!]
\caption{\label{tab:FrFy}
	Radial and vertical components of ``pure'' normal and skew S-multipoles'
	field.}
\begin{ruledtabular}
\begin{tabular}{llll}
%------------------------------------------------------------------------------%
	&
n	&							\\\hline
								\\[-0.3cm]
%------------------------------------------------------------------------------%
$\Fron$
&0	& calibration &
	$\,\,\,\,$---					\\[0.3cm]
&1	& normal dipole &
	$\quad\displaystyle\,0 \phantom{\frac{1}{1}}$	\\[0.3cm]
&2	& normal quadrupole &
	$-\displaystyle \frac{1}{1!}\frac{y}{\rho}\left\{
	\!\!\!\!\!\!\!\!\!\!\!\phantom{\left[ \frac{1}{1} \right]}
	1
	\right\} \,\overline{C}_{2}$				\\[0.3cm]
&3	& normal sextupole &
	$-\displaystyle \frac{1}{2!}\frac{y}{\rho}\left\{
	\left[2\,\frac{\rho^2-1}{2}\right]
	\right\}
	\,\overline{C}_{3}$					\\[0.3cm]
&4	& normal octupole &
	$-\displaystyle \frac{1}{3!}\frac{y}{\rho}\left\{
	\left[
		 -3\,\frac{\rho^2-1}{2}- y^2
	\right]
	+ 3\,\rho^2\ln \rho
	\right\}
	\,\overline{C}_{4}$					\\[0.3cm]
&5	& normal decapole &
	$-\displaystyle \frac{1}{4!}\frac{y}{\rho}\left\{
	\left[
		 4\,\frac{3(\rho^2+1)}{4}\frac{\rho^2-1}{2}
		-4\,\frac{\rho^2-1}{2}\,y^2
	\right]
	- 6\,\rho^2\ln \rho
	\right\}
	\,\overline{C}_{5}$					\\[-0.3cm]
								\\\hline
								\\[-0.3cm]
%------------------------------------------------------------------------------%
$\Fyon$
&0	& calibration &
	$\,\,\,\,$---						\\[0.3cm]
&1	& normal dipole &
	$\quad\!\displaystyle\frac{1}{0!}
	\left\{
		1\phantom{\frac{1}{1}}\!\!\!\!\!
	\right\}\,\overline{C}_{1}$				\\[0.3cm]
&2	& normal quadrupole &
	$\quad\!\displaystyle \frac{1}{1!}\left\{
	\!\!\!\!\!\!\!\!\!\!\!\phantom{\left[ \frac{1}{1} \right]}
		\ln \rho\phantom{\frac{1}{1}}\!\!\!\!
	\right\} \,\overline{C}_{2}$				\\[0.3cm]
&3	& normal sextupole &
	$\quad\!\displaystyle \frac{1}{2!}\left\{
	\left[
		\frac{\rho^2-1}{2}-y^2
	\right] - \ln \rho
	\right\}
	\,\overline{C}_{3}$					\\[0.3cm]
&4	& normal octupole &
	$\quad\!\displaystyle \frac{1}{3!}\left\{
	\left[
		-3\,\frac{\rho^2-1}{2}
	\right]+
	3\,\left(
		\frac{\rho^2+1}{2}-y^2
	\right)\ln \rho
	\right\}
	\,\overline{C}_{4}$					\\[0.3cm]
&5	& normal decapole &
	$\quad\!\displaystyle \frac{1}{4!}\left\{
	\left[
		 \frac{3(\rho^4+4\,\rho^2-5)}{8}
		-6\,\frac{\rho^2-1}{2}\,y^2
		+y^4
	\right]
	-3\,\left(
		\frac{1}{2}+\rho^2-2\,y^2
	\right)\,\ln \rho
	\right\}
	\,\overline{C}_{5}$					\\[-0.3cm]
								\\\hline
								\\[-0.3cm]
%------------------------------------------------------------------------------%
$\Frun$
&0	& calibration &
	$\,\,\,\,$---						\\[0.3cm]
&1	& skew dipole &
	$\quad\!\displaystyle \frac{1}{0!}\frac{1}{\rho}
	\left\{
		1 \phantom{\frac{1}{1}}\!\!\!\!
	\right\}\,\underline{C}_{\,1}$				\\[0.3cm]
&2	& skew quadrupole &
	$\quad\!\displaystyle \frac{1}{1!}\frac{1}{\rho}\left\{
	\left[ \frac{\rho^2-1}{2} \right]
	\right\}\,\underline{C}_{\,2}$				\\[0.3cm]
&3	& skew  sextupole &
	$\quad\!\displaystyle \frac{1}{2!}\frac{1}{\rho}\left\{
	\left[-\frac{\rho^2-1}{2} - y^2\right]
	+\rho^2\ln \rho
	\right\}
	\,\underline{C}_{\,3}$					\\[0.3cm]
&4	& skew octupole &
	$\quad\!\displaystyle \frac{1}{3!}\frac{1}{\rho}\left\{
	\left[
		 \frac{3(\rho^2+1)}{4}\frac{\rho^2-1}{2}
		-3\,\frac{\rho^2-1}{2}\,y^2
	\right]
	- \frac{3}{2}\,\rho^2\ln \rho
	\right\}
	\,\underline{C}_{\,4}$					\\[0.3cm]
&5	& skew decapole &
	$\quad\!\displaystyle \frac{1}{4!}\frac{1}{\rho}\left\{
	\left[
		-\frac{3(5\,\rho^4-4\,\rho^2-1)}{8}
		+6\,\frac{\rho^2-1}{2}\,y^2
		+y^4
	\right]
	+ \frac{3(2+\rho^2-4\,y^2)}{2}\,\rho^2\ln \rho
	\right\}
	\,\underline{C}_{\,5}$					\\[-0.3cm]
								\\\hline
								\\[-0.3cm]
%------------------------------------------------------------------------------%
$\Fyun$
&0	& calibration &
	$\,\,\,\,$---					\\[0.3cm]
&1	& skew dipole &
	$\quad\displaystyle\,0 \phantom{\frac{1}{1}}$	\\[0.3cm]
&2	& skew quadrupole &
	$\quad\!\displaystyle \frac{1}{1!}\,y\left\{
	\!\!\!\!\!\!\!\!\!\!\!\phantom{\left[ \frac{1}{1} \right]}
	1
	\right\}\,\underline{C}_{\,2}$				\\[0.3cm]
&3	& skew sextupole &
	$\quad\!\displaystyle \frac{1}{2!}\,y\left\{
	\!\!\!\!\!\!\!\!\!\!\!\phantom{\left[ \frac{1}{1} \right]}
		2\,\ln \rho
	\right\}
	\,\underline{C}_{\,3}$					\\[0.3cm]
&4	& skew octupole &
	$\quad\!\displaystyle \frac{1}{3!}\,y\left\{
	\left[
		3\,\frac{\rho^2-1}{2} - y^2
	\right]
	-3\,\ln \rho
	\right\}
	\,\underline{C}_{\,4}$					\\[0.3cm]
&5	& skew decapole &
	$\quad\!\displaystyle \frac{1}{4!}\,y\left\{
	\left[
		-12\,\frac{\rho^2-1}{2}
	\right]
	+4\,\left(
		3\,\frac{\rho^2+1}{2}-y^2
	\right)\ln \rho
	\right\}
	\,\underline{C}_{\,5}$
%------------------------------------------------------------------------------%
\end{tabular}
\end{ruledtabular}
\end{table*}
%------------------------------------------------------------------------------%

% F, G, G/r Taylor ============================================================%
%==============================================================================%

%-------------------------------------------------------------------------------
\begingroup
\squeezetable
%-------------------------------------------------------------------------------
\begin{table*}[t!]
\caption{\label{tab:FGTaylor}
Maclaurin series of $\Fn(x)$, $\Gn(x)$ and $\frac{\Gn(x)}{1+x}$; they are also
Taylor polynomials of $\Fn(\rho)$, $\Gn(\rho)$ and $\frac{\Gn(\rho)}{\rho}$ at 
$\rho=1$.}
\begin{ruledtabular}
\begin{tabular}{ll}
%-------------------------------------------------------------------------------
$n$	& $\mathrm{T}(\Fn)$					\\\hline
	&							\\[-0.1cm]
%-------------------------------------------------------------------------------
%-------------------------------------------------------------------------------
0	&
$\displaystyle 1 \phantom{\frac{1}{1}}$				\\[0.25cm]
%-------------------------------------------------------------------------------
1	&
$\displaystyle x
-\frac{1}{2} \,x^2
+\frac{1}{3} \,x^3
-\frac{1}{4} \,x^4
+\frac{1}{5} \,x^5
-\frac{1}{6} \,x^6
+\frac{1}{7} \,x^7
-\frac{1}{8} \,x^8
+\frac{1}{9} \,x^9
-\frac{1}{10}\,x^{10}
+O(x^{11})$							\\[0.25cm]
%-------------------------------------------------------------------------------
2	&
$\displaystyle x^2
-\frac{1}{3} \,x^3
+\frac{1}{4} \,x^4
-\frac{1}{5} \,x^5
+\frac{1}{6} \,x^6
-\frac{1}{7} \,x^7
+\frac{1}{8} \,x^8
-\frac{1}{9} \,x^9
+\frac{1}{10}\,x^{10}
-\frac{1}{11}\,x^{11}
+O(x^{12})$							\\[0.25cm]
%-------------------------------------------------------------------------------
3	&
$\displaystyle x^3
-\frac{1} {2}  \,x^4
+\frac{7} {20} \,x^5
-\frac{11}{40} \,x^6
+\frac{8} {35} \,x^7
-\frac{11}{56} \,x^8
+\frac{29}{168}\,x^9
-\frac{37}{240}\,x^{10}
+\frac{23}{165}\,x^{11}
-\frac{7} {55} \,x^{12}
+O(x^{13})$							\\[0.25cm]
%-------------------------------------------------------------------------------
4	&
$\displaystyle x^4
-\frac{2} {5}  \,x^5
+\frac{3} {10} \,x^6
-\frac{17}{70} \,x^7
+\frac{23}{112}\,x^8
-\frac{5} {28} \,x^9
+\frac{19}{120}\,x^{10}
-\frac{47}{330}\,x^{11}
+\frac{57}{440}\,x^{12}
-\frac{17}{143}\,x^{13}
+O(x^{14})$							\\[0.25cm]
%-------------------------------------------------------------------------------
5	&
$\displaystyle x^5
-\frac{1}   {2}   \,x^6
+\frac{5}   {14}  \,x^7
-\frac{2}   {7}   \,x^8
+\frac{27}  {112} \,x^9
-\frac{47}  {224} \,x^{10}
+\frac{689} {3696}\,x^{11}
-\frac{355} {2112}\,x^{12}
+\frac{263} {1716}\,x^{13}
-\frac{1129}{8008}\,x^{14}
+O(x^{15})$							\\[0.25cm]
%-------------------------------------------------------------------------------
6	&
$\displaystyle x^6
-\frac{3}   {7}    \,x^7
+\frac{9}   {28}   \,x^8
-\frac{11}  {42}   \,x^9
+\frac{25}  {112}  \,x^{10}
-\frac{241} {1232} \,x^{11}
+\frac{123} {704}  \,x^{12}
-\frac{181} {1144} \,x^{13}
+\frac{2319}{16016}\,x^{14}
-\frac{535} {4004} \,x^{15}
+O(x^{16})$							\\[0.25cm]
%-------------------------------------------------------------------------------
7	&
$\displaystyle x^7
-\frac{1}   {2}    \,x^8
+\frac{13}  {36}   \,x^9
-\frac{7}   {24}   \,x^{10}
+\frac{131} {528}  \,x^{11}
-\frac{689} {3168} \,x^{12}
+\frac{5339}{27456}\,x^{13}
-\frac{9683}{54912}\,x^{14}
+\frac{16}  {99}   \,x^{15}
-\frac{1367}{9152} \,x^{16}
+O(x^{17})$							\\[0.25cm]
%-------------------------------------------------------------------------------
8	&
$\displaystyle x^8
-\frac{4}   {9}    \,x^9
+\frac{1}   {3}    \,x^{10}
-\frac{3}   {11}   \,x^{11}
+\frac{185} {792}  \,x^{12}
-\frac{353} {1716} \,x^{13}
+\frac{1267}{6864} \,x^{14}
-\frac{3457}{20592}\,x^{15}
+\frac{5647}{36608}\,x^{16}
-\frac{855} {5984} \,x^{17}
+O(x^{18})$							\\[0.25cm]
%-------------------------------------------------------------------------------
9	&
$\displaystyle x^9
-\frac{1}     {2}      \,x^{10}
+\frac{4}     {11}     \,x^{11}
-\frac{13}    {44}     \,x^{12}
+\frac{289}   {1144}   \,x^{13}
-\frac{509}   {2288}   \,x^{14}
+\frac{457}   {2288}   \,x^{15}
-\frac{2}     {11}     \,x^{16}
+\frac{9461}  {56576}  \,x^{17}
-\frac{192991}{1244672}\,x^{18}
+O(x^{19})$							\\[0.4cm]
%-------------------------------------------------------------------------------
%-------------------------------------------------------------------------------
$n$	& $\mathrm{T}(\Gn)$					\\\hline
	&							\\[-0.1cm]
%-------------------------------------------------------------------------------
%-------------------------------------------------------------------------------
0	&
$\displaystyle 1 \phantom{\frac{1}{1}}$				\\[0.25cm]
%-------------------------------------------------------------------------------
1	&
$\displaystyle x
+\frac{1}{2}   \,x^2$							
\\[0.3cm]
%-------------------------------------------------------------------------------
2	&
$\displaystyle x^2
+\frac{1}{3}   \,x^3
-\frac{1}{12}  \,x^4
+\frac{1}{30}  \,x^5
-\frac{1}{60}  \,x^6
+\frac{1}{105} \,x^7
-\frac{1}{168} \,x^8
+\frac{1}{252} \,x^9
-\frac{1}{360} \,x^{10}
+\frac{1}{495} \,x^{11}
+O(x^{12})$							\\[0.25cm]
%-------------------------------------------------------------------------------
3	&
$\displaystyle x^3
+\frac{1}{2}   \,x^4
-\frac{1}{20}  \,x^5
+\frac{1}{40}  \,x^6
-\frac{1}{70}  \,x^7
+\frac{1}{112} \,x^8
-\frac{1}{168} \,x^9
+\frac{1}{240} \,x^{10}
-\frac{1}{330} \,x^{11}
+\frac{1}{440} \,x^{12}
+O(x^{13})$							\\[0.25cm]
%-------------------------------------------------------------------------------
4	&
$\displaystyle x^4
+\frac{2} {5}   \,x^5
-\frac{1} {10}  \,x^6
+\frac{3} {70}  \,x^7
-\frac{13}{560} \,x^8
+\frac{1} {70}  \,x^9
-\frac{1} {105} \,x^{10}
+\frac{31}{4620}\,x^{11}
-\frac{13}{2640}\,x^{12}
+\frac{8} {2145}\,x^{13}
+O(x^{14})$							\\[0.25cm]
%-------------------------------------------------------------------------------
5	&
$\displaystyle x^5
+\frac{1} {2}    \,x^6
-\frac{1} {14}   \,x^7
+\frac{1} {28}   \,x^8
-\frac{1} {48}   \,x^9
+\frac{3} {224}  \,x^{10}
-\frac{17}{1848} \,x^{11}
+\frac{7} {1056} \,x^{12}
-\frac{17}{3432} \,x^{13}
+\frac{61}{16016}\,x^{14}
+O(x^{15})$							\\[0.25cm]
%-------------------------------------------------------------------------------
6	&
$\displaystyle x^6
+\frac{3}  {7}    \,x^7
-\frac{3}  {28}   \,x^8
+\frac{1}  {21}   \,x^9
-\frac{3}  {112}  \,x^{10}
+\frac{3}  {176}  \,x^{11}
-\frac{173}{14784}\,x^{12}
+\frac{271}{32032}\,x^{13}
-\frac{37}{5824}  \,x^{14}
+\frac{59}{12012} \,x^{15}
+O(x^{16})$							\\[0.25cm]
%-------------------------------------------------------------------------------
7	&
$\displaystyle x^7
+\frac{1}  {2}    \,x^8
-\frac{1}  {12}   \,x^9
+\frac{1}  {24}   \,x^{10}
-\frac{13} {528}  \,x^{11}
+\frac{17} {1056} \,x^{12}
-\frac{103}{1952} \,x^{13}
+\frac{151}{18304}\,x^{14}
-\frac{43} {6864} \,x^{15}
+\frac{179}{36609}\,x^{16}
+O(x^{17})$							\\[0.25cm]
%-------------------------------------------------------------------------------
8	&
$\displaystyle x^8
+\frac{4}   {9}     \,x^9
-\frac{1}   {9}     \,x^{10}
+\frac{5}   {99}    \,x^{11}
-\frac{23}  {792}   \,x^{12}
+\frac{97}  {5148}  \,x^{13}
-\frac{271} {20592} \,x^{14}
+\frac{199} {20592} \,x^{15}
-\frac{2425}{329472}\,x^{16}
+\frac{1009}{175032}\,x^{17}
+O(x^{18})$							\\[0.25cm]
%-------------------------------------------------------------------------------
9	&
$\displaystyle x^9
+\frac{1}   {2}      \,x^{10}
-\frac{1}   {11}     \,x^{11}
+\frac{1}   {22}     \,x^{12}
-\frac{31}  {1144}   \,x^{13}
+\frac{41}  {2288}   \,x^{14}
-\frac{29}  {2288}   \,x^{15}
+\frac{43}  {4576}   \,x^{16}
-\frac{4489}{622336} \,x^{17}
+\frac{7079}{1244672}\,x^{18}
+O(x^{19})$							\\[0.4cm]
%-------------------------------------------------------------------------------
%-------------------------------------------------------------------------------
$n$	& $\mathrm{T}(\Gn/\rho)$				\\\hline
	&							\\[-0.1cm]
%-------------------------------------------------------------------------------
0	&
$\displaystyle 1
-	x
+	x^2
-	x^3
+	x^4
-	x^5
+	x^6
-	x^7
+	x^8
-	x^9
+O(x^{10})$							\\[0.25cm]
%-------------------------------------------------------------------------------
1	&
$\displaystyle x
-\frac{1}{2} \,x^2
+\frac{1}{2} \,x^3
-\frac{1}{2} \,x^4
+\frac{1}{2} \,x^5
-\frac{1}{2} \,x^6
+\frac{1}{2} \,x^7
-\frac{1}{2} \,x^8
+\frac{1}{2} \,x^9
-\frac{1}{2} \,x^{10}
+O(x^{11})$							\\[0.25cm]
%-------------------------------------------------------------------------------
2	&
$\displaystyle x^2
-\frac{2} {3}  \,x^3
+\frac{7} {12} \,x^4
-\frac{11}{20} \,x^5
+\frac{8} {15} \,x^6
-\frac{11}{21} \,x^7
+\frac{29}{56} \,x^8
-\frac{37}{72} \,x^9
+\frac{23}{45} \,x^{10}
-\frac{28}{55} \,x^{11}
+O(x^{12})$							\\[0.25cm]
%-------------------------------------------------------------------------------
3	&
$\displaystyle x^3
-\frac{1}  {2}  \,x^4
+\frac{9}  {20} \,x^5
-\frac{17} {40} \,x^6
+\frac{23} {56} \,x^7
-\frac{45} {112}\,x^8
+\frac{19} {48} \,x^9
-\frac{47} {120}\,x^{10}
+\frac{171}{440}\,x^{11}
-\frac{17} {44} \,x^{12}
+O(x^{13})$							\\[0.25cm]
%-------------------------------------------------------------------------------
4	&
$\displaystyle x^4
-\frac{3}   {5}   \,x^5
+\frac{1}   {2}   \,x^6
-\frac{16}  {35}  \,x^7
+\frac{243} {560} \,x^8
-\frac{47}  {112} \,x^9
+\frac{689} {1680}\,x^{10}
-\frac{71}  {176} \,x^{11}
+\frac{263} {660} \,x^{12}
-\frac{1129}{2860}\,x^{13}
+O(x^{14})$							\\[0.25cm]
%-------------------------------------------------------------------------------
5	&
$\displaystyle x^5
-\frac{1}   {2}    \,x^6
+\frac{3}   {7}    \,x^7
-\frac{11}  {28}   \,x^8
+\frac{125} {336}  \,x^9
-\frac{241} {672}  \,x^{10}
+\frac{123} {352}  \,x^{11}
-\frac{181} {528}  \,x^{12}
+\frac{773} {2288} \,x^{13}
-\frac{2675}{8008} \,x^{14}
+O(x^{15})$							\\[0.25cm]
%-------------------------------------------------------------------------------
6	&
$\displaystyle x^6
-\frac{4}   {7}    \,x^7
+\frac{13}  {28}   \,x^8
-\frac{5}   {12}   \,x^9
+\frac{131} {336}  \,x^{10}
-\frac{689} {1848} \,x^{11}
+\frac{5339}{14784}\,x^{12}
-\frac{9683}{27456}\,x^{13}
+\frac{80}  {231}  \,x^{14}
-\frac{1367}{4004} \,x^{15}
+O(x^{16})$							\\[0.25cm]
%-------------------------------------------------------------------------------
7	&
$\displaystyle x^7
-\frac{1}    {2}     \,x^8
+\frac{5}    {12}    \,x^9
-\frac{3}    {8}     \,x^{10}
+\frac{185}  {528}   \,x^{11}
-\frac{353}  {1056}  \,x^{12}
+\frac{8869} {27456} \,x^{13}
-\frac{17285}{54912} \,x^{14}
+\frac{5647} {18304} \,x^{15}
-\frac{855}  {2816}  \,x^{16}
+O(x^{17})$							\\[0.25cm]
%-------------------------------------------------------------------------------
8	&
$\displaystyle x^8
-\frac{5}     {9}     \,x^9
+\frac{4}     {9}     \,x^{10}
-\frac{13}    {33}    \,x^{11}
+\frac{289}   {792}   \,x^{12}
-\frac{3563}  {10296} \,x^{13}
+\frac{2285}  {6864}  \,x^{14}
-\frac{32}    {99}    \,x^{15}
+\frac{9461}  {29952} \,x^{16}
-\frac{192991}{622336}\,x^{17}
+O(x^{18})$							\\[0.25cm]
%-------------------------------------------------------------------------------
9	&
$\displaystyle x^9
-\frac{1}     {2}      \,x^{10}
+\frac{9}     {22}     \,x^{11}
-\frac{4}     {11}     \,x^{12}
+\frac{35}    {104}    \,x^{13}
-\frac{729}   {2288}   \,x^{14}
+\frac{175}   {572}    \,x^{15}
-\frac{1357}  {4576}   \,x^{16}
+\frac{13851} {47872}  \,x^{17}
-\frac{353047}{1244672}\,x^{18}
+O(x^{19})$
%-------------------------------------------------------------------------------
\end{tabular}
\end{ruledtabular}
\end{table*}
%-------------------------------------------------------------------------------
\endgroup
%-------------------------------------------------------------------------------

% The \nocite command causes all entries in a bibliography to be printed out
% whether or not they are actually referenced in the text. This is appropriate
% for the sample file to show the different styles of references, but authors
% most likely will not want to use it.
\nocite{*}

\bibliography{bibfile}% Produces the bibliography via BibTeX.

\end{document}